\documentclass[12pt]{article}
\pdfoutput=1
\usepackage{epsf,amsfonts,amssymb,epsfig,amsmath,mathtools,graphics,slashed}
\usepackage{hyperref,hep,graphicx,subfig}
\usepackage{rotating}
\newcommand{\x}{\tilde x}
\newcommand{\y}{\tilde y}

\newcommand{\nn}{\notag \\}

\makeatletter

\@addtoreset{equation}{section}
\makeatother

\begin{document}

\begin{titlepage}

\vfill

\begin{flushright}
DCPT-19/31\\
Imperial/TP/2020/JG/01
\end{flushright}

\vfill

\begin{center}
   \baselineskip=16pt
   {\Large\bf Holographic Abrikosov Lattices}
 \vskip 1.5cm
      Aristomenis Donos$^1$, Jerome P. Gauntlett$^2$ and Christiana Pantelidou$^{1,3}$\\
   \vskip .6cm
    \vskip .6cm
      \begin{small}
      \textit{$^1$ Centre for Particle Theory and Department of Mathematical Sciences,\\ Durham University,
       Durham, DH1 3LE, U.K.}
        \end{small}\\
      \vskip .6cm
      \begin{small}
      \textit{$^2$Blackett Laboratory, 
        Imperial College\\ London, SW7 2AZ, U.K.}
        \end{small}\\
      \vskip .6cm
      \begin{small}
      \textit{$^3$       School of Mathematics, Trinity College Dublin, Dublin 2, Ireland}
        \end{small}\\*[.6cm]
         
\end{center}

%\vfill

\begin{center}
\textbf{Abstract}
\end{center}
\begin{quote}
We study black hole solutions of $D=4$ Einstein-Maxwell theory coupled to a charged scalar field that
are holographically dual to a $d=3$ conformal field theory with a non-vanishing chemical potential and
constant magnetic field. We numerically construct black hole solutions that are dual to a superfluid phase
with a periodic lattice of vortices. 
For the specific model we investigate, we find that
the thermodynamically preferred configuration is given by a triangular lattice 
and moreover the vortices are associated with the lowest Landau level. We also construct black holes
describing a lattice of vortices associated with the next to lowest Landau level and while these are not thermodynamically preferred
they exhibit some interesting features that could be realised for other holographic models.
\end{quote}

    \vskip .6cm

\center{\it Dedicated to the memory of Steven Gubser}
\vfill
\end{titlepage}

\setcounter{equation}{0}
\tableofcontents
\newpage

\section{Introduction}

Holography provides a controlled theoretical framework to study strongly coupled quantum field theories.
In seeking possible applications to real systems an important development, pioneered by Steven Gubser, was the realisation that holographic matter can exist in a superfluid phase \cite{Gubser:2008px}.
In the simplest set-up one considers Einstein-Maxwell theory coupled to a charged scalar field with an $AdS$ vacuum solution that is dual to a conformal field theory with a global abelian symmetry. 
When the CFT is held at finite chemical potential,
the unbroken phase at high temperature
is described by an electrically charged, planar AdS-RN black hole solution with
vanishing charged scalar field. This black hole is unstable below some critical temperature and the system condenses into a superfluid phase which is described by an electrically charged black hole carrying a halo of charged 
scalar hair \cite{Gubser:2008px, Hartnoll:2008vx,Hartnoll:2008kx}.

In this paper we study superfluid phases of holographic matter held at finite chemical potential, with the addition of an external magnetic field. Over the past ten years this topic has been studied from several different points of view. Switching on the magnetic field suppresses the superfluid phase transition, as one might expect\footnote{In this paper we only study $s$-wave superfluids. For holographic $p$-wave superfluids, it has been shown that
magnetic fields can induce a superfluid instability at vanishing chemical potential \cite{Ammon:2011je}. Furthermore, holographic vortex lattices for $p$-wave superfluids, in a probe approximation, have been studied at finite chemical potential in [5] and for non-zero magnetic fields in [6]. Both of these constructions are perturbatively close to the phase transition point.
}. Indeed, and as we will review, the critical temperature at which the superfluid instability sets in 
decreases as one increases the magnetic field and for large enough magnetic field the instability is longer present 
\cite{Nakano:2008xc,Albash:2008eh,Hartnoll:2008kx}. Below the critical temperature one expects the existence of
vortices. The defining feature of a vortex is that the phase of the complex field has non-zero winding as one goes around the vortex and, as a consequence, the complex field vanishes at the core of the vortex. 
Using certain probe approximations, where there is no back-reaction on the metric, 
constructions of vortex-like solutions were made in \cite{Albash:2009ix,Albash:2009iq,Montull:2009fe}. 
A further development, again in a probe approximation, was the construction of a vortex lattice\footnote{Holographic vortices associated with a rotating superfluid on a disc were discussed in a probe approximation in
\cite{Xia:2019eje,Yang:2019ibe}, by imposing non-standard boundary conditions on the disc throughout the bulk.}, 
using vortices in the lowest Landau level \cite{Maeda:2009vf}. 
Going beyond the probe approximation the existence of 
a vortex lattice solution was argued for in \cite{Bao:2013fda}, again in the lowest Landau level, by considering a perturbative expansion about a purely magnetic $AdS_2\times\mathbb{R}^2$ zero temperature ground state. 
An approximate back-reacted vortex lattice solution was recently constructed in \cite{Tallarita:2019amp} after imposing by hand a certain circular symmetry on each unit cell.

The purpose of this paper is to report on the first numerical construction of fully back-reacted black hole solutions, without any approximations, that describe
a periodic vortex lattice in a superfluid phase. {\it A priori} it is not clear what the thermodynamically preferred shape of the vortex lattice will be. If one is just below the critical temperature and one is able to utilise a Landau-Ginzburg description, one finds that a triangular lattice associated with the lowest Landau level is preferred (for a review see \cite{RevModPhys.82.109}). Thus, one might expect that in holography the triangular vortex lattice is also the preferred configuration at least just below the critical temperature, and in the
context of the probe approximation some arguments supporting this conclusion were given in \cite{Maeda:2009vf}.
However, it is worth emphasising that in certain holographic situations the Landau-Ginzburg does not effectively capture the properties of the phase transitions near the critical temperature \cite{Banks:2015aca} and so this conclusion may not be valid in general. In any event, as the temperature is lowered the Landau-Ginzburg description becomes less useful and it is no longer clear what shape the preferred lattice will take. In fact various different shapes are realised in real superconductors as well as transitions to other phases such as vortex liquids and glasses \cite{RevModPhys.82.109}. It is therefore of significant interest to find out what can happen in the context of holography and this paper is a step in exploring what is possible.

The $D=4$ gravitational model that we will consider 
couples the metric to a Maxwell field and a complex scalar field. The model
has an $AdS_4$ vacuum solution, with vanishing Maxwell and scalar field, which is dual to the underlying $d=3$ CFT that we want to
study both with non-vanishing magnetic field and 
at finite chemical with respect to the global $U(1)$ symmetry. 
It also has another $AdS_4$ solution
with non-vanishing scalar field which describes the IR behaviour of the superfluid phase with vanishing magnetic field, $B=0$. When $B\ne 0$ the
high temperature, unbroken phase is described by the dyonic AdS-RN black hole solution. We review the linearised instabilities of this
black hole and, show that the critical temperature at which it becomes unstable just depends on the Landau level of the linearised perturbation, with the 
lowest Landau level having the highest critical temperature. Within the linearised framework one can then
construct vortex lattice solutions, parametrised by the Landau level as well as two additional parameters which determine the shape of the lattice. 

For a specific value of the magnetic field, we construct the back-reacted vortex lattice associated with the lowest Landau level. By minimising the free energy of the black hole solutions with respect to the remaining two shape parameters, we show that the triangular lattice is the preferred configuration
for the specific temperatures we consider. 
While a further refinement of our numerics is required in order to construct black holes at very low temperatures, 
at the end of the paper we discuss a plausible zero temperature ground state solution.
We also construct vortex lattice solutions for the second lowest Landau level,
which appear at a critical temperature that is lower than those associated with the lowest Landau level.
Interestingly, the thermodynamically preferred black holes in this class are associated with
infinitely thin and long lattice structures, indicating the existence of an interesting kind of linear vortex defect\footnote{Our results near the critical temperature, in particular,
strongly indicate that these linear structures should also appear within the Landau-Ginzburg framework and it would be interesting to directly confirm this.}
For temperatures when both black holes exist we find that
the triangular vortex lattice associated with the lowest Landau level is always thermodynamically
preferred. It is worth noting, however, that this conclusion certainly depends on the bulk gravitational model, a point we return to in the discussion section.

The holographic black hole solutions that we construct consist of a lattice of vortex tubes
that stretch out from the black hole horizon and extend to the asymptotic boundary. Since the proper radius of the vortex tubes grows as one approaches the boundary they have a funnel-type structure. We emphasise that in the boundary theory the Maxwell field is not dynamical and hence the magnetic field is not localised inside the vortices. Instead, the vortices are associated with circulating currents in the boundary theory. It should be noted that this set-up is different from the
usual superfluid vortices which carry quantised orbital angular momentum.
It would be interesting to know if there are experimental setups
where the configurations we discuss can be realised.

There are various other constructions of fully back-reacted black holes describing spatially modulated phases, in
which translations are spontaneously broken including \cite{Donos:2012gg,Donos:2012wi,Donos:2013wia,Withers:2013loa,Withers:2013kva,Rozali:2013ama,Donos:2013woa,Withers:2014sja,Donos:2015eew,Donos:2016hsd}. In particular, our construction shares several similarities with the work of \cite{Donos:2015eew}.

The plan of the rest of this paper is as follows.
In section \ref{sec:setup} we introduce the bottom-up holographic model of interest.
In section \ref{sec:BHs} we discuss various details of the gravitational  boundary value problem relevant to the formation of an Abrikosov lattice of vortices. We then discuss the numerical techniques we use to construct the broken phase black hole solutions along with their thermodynamics. In section \ref{sec:Results} we discuss the main 
results of the numerical analysis and we conclude with some discussion in section \ref{disc}. Appendix \ref{appendixA} has some details about the asymptotic expansions and one point functions while appendix \ref{appendixB} contains some comments
about our numerical scheme and convergence properties.

\section{Set-up}\label{sec:setup}
We will consider a bulk action in $D=4$ spacetime dimensions of the form
\begin{align}\label{eq:action}
S&=
%\frac{1}{16\pi\,G}\,
\int\,d^{4}x\,\sqrt{-g}\,\Big( R-V-\frac{1}{2}D_{\mu}\psi\,D^{\mu}\bar{\psi}-\frac{1}{4}F^{2}\Big)\,,%\nn
%F&=dA,\quad D_{\mu}\psi=\nabla_{\mu}\psi+i\,q\,A_{\mu}\,\psi\,.
\end{align}
with $F=dA$ and $D_{\mu}\psi=\nabla_{\mu}\psi+i\,q\,A_{\mu}\,\psi$. We also take
$V=V(|\psi|^2)$ so the action is invariant under the gauge transformation $\psi\to e^{-iq\Lambda}\psi$, $A\to A+d\Lambda$.
For simplicity we have fixed Newton's constant so that $16\pi G=1$.
The equations of motion associated with \eqref{eq:action} are given by
\begin{align}\label{eq:eom}
R_{\mu\nu}-\frac{1}{2}D_{\left(\mu\right.}\psi\,D_{\left.\nu\right)}\bar{\psi}-\frac{1}{2}V\,g_{\mu\nu}+\frac{1}{2}\Big(\frac{1}{4}\,g_{\mu\nu}F_{\lambda\rho}F^{\lambda\rho}-F_{\mu\rho}F_{\nu}{}^{\rho} \Big)&=0\,,\nn
\nabla_{\mu}F^{\mu\nu}+i\frac{q}{2}\left( \bar{\psi}\,D^{\nu}\psi-\psi\,D^{\nu}\bar{\psi}\right)&=0\,,\nn
D_{\mu}D^{\mu}\psi-2\,V^{\prime}\psi&=0\,.
\end{align}

We focus on the specific choice of potential given by
\begin{equation}\label{potchoice}
V=-6 +\frac{m^2}{2}\left| \psi\right|^{2}+\frac{1}{2}\left| \psi\right|^{4}\,,\qquad m^2=-2\,.
%G\left( \left| \psi\right|^{2}\right)=1\,,\quad Z\left( \left| \psi\right|^{2}\right)=1\,.
\end{equation}
The equations of motion then admit a unit-radius $AdS_4$ vacuum solution with $A=\psi=0$, which is 
dual to a $d=3$ CFT with an abelian global symmetry. We will choose boundary conditions so that
the complex scalar field $\psi$, of charge $q$, is dual to an operator 
$\mathcal{O}_\psi$, with scaling dimension $\Delta_\psi=2$.
Our choice of potential is such that there is another $AdS_4$ solution with
$|\psi|=1$ and radius squared  equal to 12/13, which also plays an important role.

We want to analyse the vacuum CFT (with $A=\psi=0$)
at finite temperature $T$, with constant chemical potential $\mu$, and constant magnetic field $B$.
The high temperature, spatially homogeneous and isotropic phase is described by the planar, dyonic AdS-Reissner-Nordstr\"om 
(AdS-RN) black brane solution. For later convenience it will be useful to write this in the following non-standard form 
\begin{align}\label{eq:norm_ansatz}
ds^{2}&=r_+^2g^{-2}(r)\,\left[-f(r)\,dt^{2}+r_+^{-2}g^{\prime}{}^{2}(r)\,f^{-1}(r)\,dr^{2}+dx^{2}+dy^{2}\right]\,,\notag\\
A&=a_{t}\,dt-B\,y\,dx,\qquad \psi=0\,,
\end{align}
with
\begin{align}\label{eq:RN}
f(r)&=\frac{1}{4r_{+}^{4}}\,\left(1-r\right)^{2}\,\left[ \left(\mu^{2}\,r_{+}^{2}+B^{2}\right)\,\left( r-2\right)^{3}r^{3}+4r_{+}^{4}\,
\left(1+2\,r+3\,r^{2}-4\,r^{3}+r^{4}\right) \right]\,,\notag\\
g&=(1-(1-r)^{2})\,,\nn
a_{t}(r)&=\mu\,\left( 1-r\right)^{2}\,.
\end{align}
The $AdS_4$ boundary is located at $r\to 0$ with the metric approaching
\begin{align}\label{rpasympads}
ds^{2}&\to \frac{dr^2}{r^2}+\frac{r_+^2}{4r^2}[-dt^{2}+dx^{2}+dy^{2}]\,,
\end{align}
and note that to get the standard form we can scale $r$ by $r_+/2$. 
The black hole horizon is located at $r=1$ in these coordinates and the 
temperature of the black hole is given by 
\begin{align}\label{tempexp}
T = \frac{1}{16\pi r_+}(12r_+^2-\mu^2-B^2/r_+^2)\,.
\end{align}
As $T\to0$ the black holes approach 
a dyonic $AdS_2\times R^2$ solution in the IR with finite entropy density.

When $B=0$, with $\Delta_\psi=2$ and any value of $q$ \cite{Denef:2009tp},
it is well known that below some critical temperature the AdS-RN solution is unstable and
the thermodynamically preferred black hole, describing a superfluid phase, has non-vanishing charged scalar hair.
The $T=0$ limit of these superfluid black hole solutions
are domain walls interpolating between the $AdS_4$ vacuum in the UV (with $A=\psi=0$)
and the second $AdS_4$ solution (with $|\psi|=1$) in the IR.
The entropy density of these black holes goes to zero as $T\to 0$, with
$s\propto T^2$.

When $B\ne 0$, the AdS-RN black hole continues to be unstable below some critical temperature that depends on 
$|qB|/\mu^2$,
up to some critical value of the magnetic field $|qB_c|/\mu^2$, as we will review shortly. Below this critical temperature it
is known that a vortex lattice can form. We will numerically construct such vortex lattice black hole solutions in the sequel.

\section{Abrikosov Lattice Black Hole Solutions}\label{sec:BHs}
\subsection{General considerations}
We will construct static black hole solutions with planar horizons that asymptote to $AdS_4$ in the UV with quasi-periodic boundary conditions for the $x$ and $y$ directions. We will use 
coordinates $(t,r,x,y)$ which are globally defined outside the black hole horizon.
At fixed $t,r$ the $x,y$ coordinates parametrise a two-torus associated with a flat metric $dx^2+dy^2$, with the following identifications
\begin{align}\label{eq:torus}
\left(x,y\right)\sim\left(x+L_{x},y\right),\quad \left(x,y\right)\sim\left(x+v\,L_{y},y+L_{y}\right)\,.
\end{align}
The parameter $v\in[0,\infty)$, which can be exchanged for an angle $\beta$ via $\cos\beta={v}/(1+v^2)^{1/2}$,
governs the deviation of the shape of the torus from a rectangular torus.
Without loss of generality, we can split the gauge field as
\begin{equation}\label{eq:gauge_split}
A=a-B \,y dx\,,
\end{equation}
where $B$ is the external magnetic field and $a=a_\mu (r,x,y)dx^\mu$ is a one-form that is globally defined on the spatial torus. 
Similarly, the metric components $g_{\mu\nu}$ are also globally defined on the torus. We now turn our attention to the complex scalar field $\psi$. It is clear from the gauge field decomposition \eqref{eq:gauge_split} that imposing $\psi$ to be well defined
on the torus would fail to satisfy the equations of motion \eqref{eq:eom}. Therefore, we take
\begin{align}\label{eq:psi_bc1}
\psi\left(r,x+L_{x},y\right)&=e^{ig_{1}\left(r,x,y\right)}\,\psi\left(r,x,y\right),\nn
\psi\left(r,x+v\,L_{y},y+L_{y}\right)&=e^{ig_{2}\left(r,x,y\right)}\,\psi\left(r,x,y\right)\,,
\end{align}
with $g_1$ and $g_2$ real functions on the torus and demand that
\begin{align}\label{eq:psi_bc2}
D_{\mu}\psi\left(r,x+L_{x},y\right)&=e^{ig_{1}\left(r,x,y\right)}\,D_{\mu}\psi\left(r,x,y\right)\,,\nonumber\\ 
D_{\mu}\,\psi\left(r,x+v\,L_{y},y+L_{y}\right)&=e^{ig_{2}\left(r,x,y\right)}\,D_{\mu}\psi\left(r,x,y\right)\,.
\end{align}
By explicit evaluation, after using \eqref{eq:gauge_split} and \eqref{eq:psi_bc1}, we have
\begin{align}\label{eq:bc_conditions}
D_{\mu}\psi\left(r,x+L_{x},y\right)&=e^{ig_{1}\left(r,x,y\right)}\,\left(D_{\mu}\psi\left(r,x,y\right)+i\,\psi \,\partial_{\mu}g_{1} \right)\,,\nn
D_{\mu}\psi\left(r,x+v\,L_{y},y+L_{y}\right)&=e^{ig_{2}\left(r,x,y\right)}\,\left(D_{\mu}\psi\left(r,x,y\right)+i\,\psi \,\left(\partial_{\mu}g_{2}-\delta^{x}_{\mu}\,qBL_{y}\right) \right)\,.
\end{align}
The general solution for the compatibility of the conditions \eqref{eq:psi_bc2} and \eqref{eq:bc_conditions} is
\begin{align}
g_{1}=c_{1},\qquad g_{2}=qBL_{y}\,x+c_{2}\,,
\end{align}
where $c_{1}$ and $c_{2}$ are real constants of integration. Thus, the boundary conditions for the complex scalar are given by
\begin{align}\label{eq:psi_bc}
\psi\left(r,x+L_{x},y\right)&=e^{ic_{1}}\,\psi\left(r,x,y\right)\,,\nn
\quad \psi\left(r,x+v\,L_{y},y+L_{y}\right)&=e^{i\left(qBL_{y}\,x+c_{2} \right)}\,\psi\left(r,x,y\right)\,.
\end{align}
Furthermore, compatibility between these two gives the quantisation condition
\begin{align}\label{eq:quant_condition}
qBL_{x}L_{y}=2\pi n\,,
\end{align}
where $n$ is an integer. 
The condition \eqref{eq:psi_bc} implies that the modulus of the complex field is periodic on the torus, while the 
phase of the complex field has winding number $n$ as we go anti-clockwise around the unit cell. Thus, each unit cell contains $n$ vortices.

For a given such solution to the equations of motion, $\psi(r,x,y)$,
we can construct another quasi-periodic solution with complex scalar field, $\tilde\psi(r,x,y)$, parametrised by two constants $x_0,y_0$,
by using the following combination of a translation and a gauge transformation
\begin{align}\label{shiftbcs}
\tilde\psi(r,x,y)=e^{-iqBy_0 x}\psi(r,x+x_0,y+y_0)\,,
\end{align}
and noting that $A(x,y)=A(x+x_0,y+y_0)+d(By_0 x)$. The boundary conditions for $\tilde\psi(x,y)$
are as in \eqref{eq:psi_bc} but with $c_1\to c_1-qBL_xy_0$ and $c_2\to c_2-qBL_y(vy_0-x_0)$.
The two constants, $c_1$ and $c_2$ can be interpreted as parametrising two Goldstone modes in the boundary theory. These Goldstone modes are associated with translations that are intertwined with the internal abelian global symmetry, due the presence of the magnetic field.
An additional Goldstone mode, associated with the abelian global symmetry is obtained by multiplying the solution by a constant phase. All of these Goldstone modes need to be fixed in order to
find a solution numerically.

\subsection{Perturbative zero modes}
We now consider zero modes around the dyonic Reissner-Nordstr\"om black hole
solution given in \eqref{eq:norm_ansatz}. 
The linearised complex scalar field equation of motion in this background is
\begin{align}\label{eq:static_eq}
%\nabla^{2}\psi+2i\,q\,A^{\mu}\,\nabla_{\mu}\psi-q^{2}\,A^{2}\,\psi+i q\,(\nabla_{\mu}A^{\mu})\,\psi-m^{2}\psi-\frac{\zeta_{0}}{2}\,F^{2}\,\psi&=0\Rightarrow\nn
\frac{g^{4}}{g^{\prime}}\,\partial_{r}\left(\frac{g^{-2}}{g^{\prime}}\,f\,\partial_{r}\psi\right)+g^{2}\,\mathcal{D}\psi+\frac{g^{2}}{f}\,q^{2}a_{t}^{2}\,\psi-m^{2}\,\psi&=0\,,
\end{align}
where
\begin{align}\label{eq:D_operator}
\mathcal{D}\psi=\left( \partial_{x}^{2}+\partial_{y}^{2}-2i\,q\,B\,y\,\partial_{x}-q^{2}B^{2}y^{2}\right)\psi\,.
\end{align}
Separating variables by writing
\newcommand{\rc}{\rho}
$\psi=\Phi(x,y)\rc(r)$, we deduce that $\Phi$ has to satisfy the eigenvalue equation
\begin{align}\label{eq:eig_equation}
\mathcal{D}\Phi=\lambda\,\Phi\,,
\end{align}
and the radial function will satisfy
\begin{align}\label{eq:z_equation}
\frac{g^{4}}{g^{\prime}}\,\partial_{r}\left(\frac{g^{-2}}{g^{\prime}}\,f\,\partial_{r}\right)+\left(g^{2}\,\lambda\,+\frac{g^{2}}{f}\,q^{2}a_{t}^{2}-m^{2}\right)\rc=0\,.
\end{align}

The task is to solve the eigenvalue equation \eqref{eq:eig_equation} with the boundary conditions
\eqref{eq:psi_bc}:
\begin{align}\label{eq:phi_bc}
\Phi\left(x+L_{x},y\right)=e^{ic_{1}}\Phi\left(x,y\right),\quad \Phi\left(x+v\,L_{y},y+L_{y}\right)=e^{ic_{2}}\,e^{i\frac{2\pi n}{L_{x}}\,x}\,\Phi\left(x,y\right)\,,
\end{align}
along with the constraint \eqref{eq:quant_condition}.
Focussing on the $x$ argument, without loss of generality and consistent with the first boundary condition,
we can write
\begin{align}
\Phi^{(\lambda)}(x,y)=e^{i\frac{c_{1}}{L_{x}}x}\,\sum_{l=-\infty}^\infty\,e^{i \frac{2l\pi}{L_{x}}\,x}\,W^{(\lambda)}_{l}(y)\,.
\end{align}
Substituting this into \eqref{eq:eig_equation}
we discover that $W^{(\lambda)}_{l}(y)$ satisfy
\begin{align}
\left[\partial_{y}^{2}-(qB)^2\left(y-\frac{2\pi\,l+c_1}{qBL_{x}} \right)^{2}-\lambda \right]W^{(\lambda)}_{l}(y)=0\,.
\end{align}
This is essentially the simple harmonic oscillator; the eigenvalues, $\lambda_j$, are labelled by an integer $j=0,1,2,\dots$, the Landau level, with
\begin{align}
W_{l}^{(\lambda_j)}(y)&=d_{l}\,\psi_{j}\left(y-\frac{2\pi\,l+c_1}{qBL_{x}} \right)\,,
\qquad \lambda_{j}=-|qB|\,\left(2j+1\right)\,,
\end{align}
and $\psi_j$ are the standard harmonic oscillator wave functions and $d_l$ are constants. 
Thus, for a given Landau level $j$, we have
 \begin{align}
\Phi^{(\lambda_j)}(x,y)=e^{i\frac{c_{1}}{L_{x}}\,x}\,\sum_{l=-\infty}^\infty\,d_{l}\,e^{i \frac{2l\pi}{L_{x}}\,x}\,\psi_{j}\left(y-\frac{2\pi\,l+c_1}{qBL_{x}} \right)\,.
\end{align}
We also need to impose the second boundary condition in \eqref{eq:phi_bc} which leads to a recursion relation on the
$d_l$ which, using the quantisation condition \eqref{eq:quant_condition}, we can solve as
\begin{align}
d_{l}=c\,e^{-i \frac{\pi l^{2}\,v L_{y}}{nL_{x}}}\,e^{\frac{il\,z}{n}}\,,\qquad
c_{2}=(c_1+\pi n)\frac{\,v\,L_{y}}{L_{x}}+z\,,
%L_{y}L_{x}&=\frac{2\pi n}{q B}\,.
\end{align}
where $z$ is a real constant and $c$ is a complex constant. 
The constant $c$ is just an overall free constant associated with the fact that we are solving
a linear equation and so we set $c=1$.
Now shifting the $y$ coordinate via $y\to y+c_1/(qBL_x)$ combined with the
gauge transformation with parameter $\Lambda=c_1/(qL_x)x$, as in \eqref{shiftbcs}, allows us to set $c_1=0$. Having set $c_1=0$, it will be convenient to shift the $x$ coordinate via $x\to x-L_x z/(2\pi n)$, again as in \eqref{shiftbcs}, in order to set $z=0$.

Thus, in summary we can write the spatial part of our zero mode as
\begin{align}\label{finalzmds}
\Phi^{(\lambda_j)}(x,y)=\sum_{l=-\infty}^\infty\,e^{-i \frac{2\pi v l^{2}}{L_{x}^2qB}}\,e^{i \frac{2\pi l}{L_{x}}\,x}\,\psi_{j}\left(y-\frac{2\pi\,l}{qBL_{x}} \right)\,,
\end{align}
which satisfies the boundary conditions given in \eqref{eq:psi_bc} with $c_1=0$ and 
$c_2=\frac{1}{2}vL^2_y qB$. We note that we have fixed all of the Goldstone modes.
For a given magnetic field, and hence fixed $qB$, the free parameters specifying these zero modes are
the torus parameters $(L_x,L_y,v)$, constrained via the quantisation condition $qBL_{x}L_{y}=2\pi n$, as well as
the Landau level $j$. The eigenvalue $\lambda$ in \eqref{eq:eig_equation} only
depends on $j$ and thus, the radial equation \eqref{eq:z_equation}, and hence the critical temperature $T_c$
at which the Reissner-Nordstr\"om solution becomes unstable, also only depend on $j$. 

\begin{figure}[h!]
\centering
\includegraphics[width=0.55\linewidth]{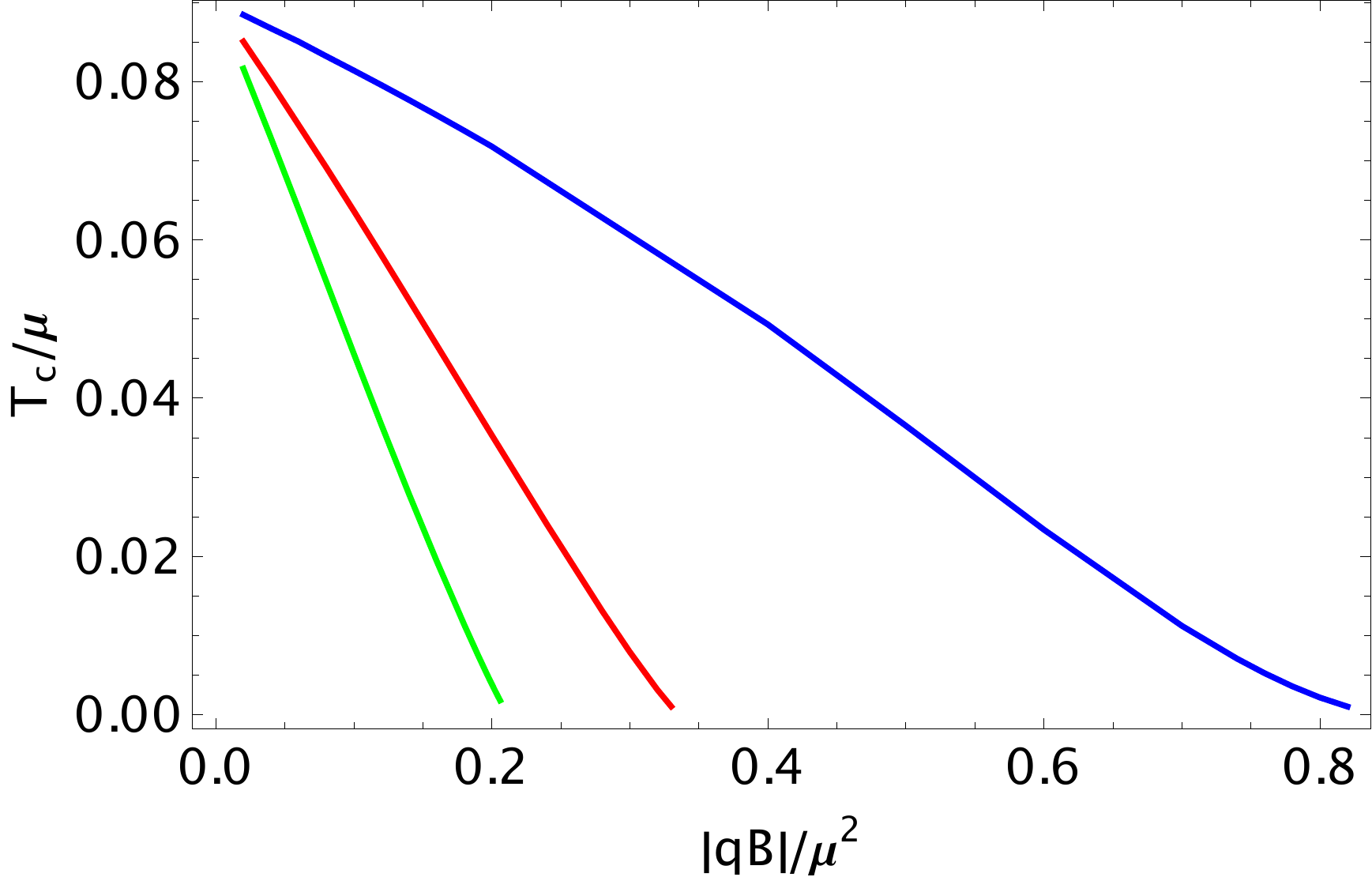}
\caption{Instabilities of the dyonic Reissner-Nordstr\"om black hole. We have plotted the 
critical temperature, $T_c$, against the magnetic field, $B$, using dimensionless units, for three different Landau levels: 
$j=0$ (blue), $j=1$ (red) and $j=2$ (green). We construct fully back-reacted black hole solutions for
$|qB|/\mu^2=0.02$ with $j=0$ and $j=1$ that first appear for values of $T_c/\mu\sim 0.088$ and $T_c/\mu\sim 0.085$, respectively.
}
\label{fig:Tc}
\end{figure}

In figure \ref{fig:Tc} we have plotted $T_c$ as a function of $|qB|$.
When $B=0$ we see that the dyonic Reissner-Nordstr\"om black hole becomes unstable to forming a superconducting state
at $T_c/\mu\sim 0.090$. The instability persists in the range $0<|qB|/\mu^2\lesssim 0.846$. For small enough $|qB|/\mu^2$
there is a finite sequence of instabilities appearing at lower temperatures that are
associated with higher Landau levels. In the sequel for $q=2$ and 
$|qB|/\mu^2=0.02$ we will construct fully back-reacted black hole solutions 
for the lowest Landau level with $j=0$, which first appear at $T_c/\mu\sim 0.088$, 
and also the first Landau level with $j=1$, which first appear at $T_c/\mu\sim 0.085$.
In each class of solutions the moduli space of solutions is parametrised by $(L_x,L_y,v)$, constrained via the quantisation condition 
$qBL_{x}L_{y}=2\pi n$. We will determine the thermodynamically preferred configurations both within each class of solutions and also show, for temperatures when they both exist, that the solutions in the lowest Landau level are preferred.

\subsection{New coordinates on the torus}
For numerical convenience, it is convenient to bring the identifications of the coordinates on the torus into the following form 
\begin{align}\label{tildeidents}
\left(\tilde x,\tilde y \right)\sim \left(\tilde x+1,\tilde y \right) \sim \left(\tilde x,\tilde y+1 \right)\,.
\end{align}
This is achieved by performing the following coordinate transformation 
\begin{align}\label{tilcchge}
\tilde x(x,y)=L_x^{-1}\,\left(x-v\,y\right),\quad \tilde y(x,y)=L_y^{-1}\,y\,.
\end{align}
In these coordinates the parameters of the torus get encoded in the metric. Indeed,
the full boundary metric and gauge field have the form
\begin{align}\label{eq:bmetric_num}
ds_{3}^{2}&=-dt^{2}+L_{x}^{2}\,d\tilde x^{2}+L_{y}^{2}(1+v^2)\,d\tilde y^{2} +2\,v L_{x}L_{y}\,d\tilde x\,d\tilde y\,,\nn
A_{\partial}&=\mu dt-BL_{x}L_{y}\,\tilde y\,d\tilde x\,- d\Lambda\,,
\end{align}
with $\Lambda(\tilde{y})=\frac{1}{2}B\, v\, L_y^2\,\tilde{y}^{2}$. The last term in 
the gauge field can be removed via a gauge transformation to give
\begin{align}\label{eq:bmetric_num2}
A_{\partial}&\to \tilde{A}_\partial=A_{\partial}+d \Lambda=\mu dt-BL_x L_y\,\tilde{y}\,d\tilde{x}\,\nn
\psi(r,\tilde{x},\tilde{y})&\to \tilde\psi=e^{-iq\,\Lambda(\tilde{y})}\,\psi(r,x(\tilde{x},\tilde{y}),y(\tilde{x},\tilde{y}))\,.
\end{align}
Furthermore, if the complex scalar field $\psi$ satisfies the quasi-periodic boundary conditions \eqref{eq:quant_condition}
with $c_1=0$ and $c_2=\frac{1}{2} v q B L_y^2/\sqrt{1+v^2}$, precisely as we chose for the zero modes given in \eqref{finalzmds},
then we have
\begin{align}\label{eq:bcscalars}
&\tilde\psi\left(r,\tilde x+1,\tilde y\right)=\tilde\psi\left(r,\tilde x,\tilde y\right),\quad \tilde\psi\left(r,\tilde  x,\tilde  y+1\right)=e^{iqBL_{x}L_{y}\,\tilde  x}\,\tilde\psi\left(r,\tilde  x,\tilde y\right)\,.
\end{align}
These are the boundary conditions we will use in the ansatz for the numerical 
integration. This fixes the
two translational Goldstone modes mentioned below \eqref{shiftbcs}. The Goldstone mode 
associated with multiplication by an overall phase is fixed by a boundary condition for the complex scalar
on the horizon as discussed in the next subsection.
Somewhat for historical reasons, in the numerical integration we
used the following rescaled quantities 
\begin{align}
\tilde L_x=(1+v^2)^{-1/4}L_x\,,\qquad 
\tilde L_y=(1+v^2)^{1/4}L_y\,,
\end{align}
in terms of which the torus part of the boundary metric in \eqref{eq:bmetric_num} takes the more symmetric form
\begin{align}\label{aasymmettwo}
ds_{3}^{2}&=-dt^{2}+(1+v^2)^{1/2}[\tilde L_{x}^{2}\,d\tilde x^{2}+\tilde L_{y}^{2}\,d\tilde y^{2}] +2\,v \tilde L_{x}\tilde L_{y}\,d\tilde x\,d\tilde y\,.
\end{align}
Note that since $L_x L_y=\tilde L_{x}\tilde L_{y}$ the form of the quantisation condition is unchanged: 
\begin{align}\label{qconag}
&qB\tilde L_{x}\tilde L_{y}=2\pi n\,.
\end{align}
For later convenience we also define a quantity $k$ such that
\begin{equation}\label{kdef}
\tilde L_x=\frac{1}{k}\sqrt{\frac{2\pi \,n}{qB}}\,,\quad \tilde L_y=k \sqrt{\frac{2\pi\, n}{qB}}\,,
\end{equation}
which respects the quantisation condition. 

Note that for the perturbative modes the solution \eqref{finalzmds} is independent of $n$. The only appearance of $n$
is in the quantisation condition \eqref{qconag}. Thus, one can obtain different values of $n$ just by rescaling 
$\tilde L_{x}$ and $\tilde L_{y}$. Going beyond the probe approximation, as we discuss in the next section, the solutions
to the non-linear equations that we have found, all have $n=1$. We do not know if it possible to
construct other solutions with $n\ne 1$.

\subsection{Numerical integration}\label{sec:numerics}
Motivated by the form of the boundary metric \eqref{eq:bmetric_num}, we consider the following ansatz
for the vortex lattice black holes
\begin{align}\label{ansatzpde}
ds^2=&\ r_+^{2}g(r)^{-2}\Big[-f(r) Q_{tt} (\eta^t)^2+\frac{g'(r)^2}{f(r)}\frac{Q_{rr}}{r_+^2} dr^2\nn
&\qquad\qquad+Q\,\sqrt{1+R^2} \left(W \tilde L_x^2 (\eta^{\x})^2+{ W}^{-1} \tilde L_y^2 (\eta^{\y})^2\right)
+2 Q R \,\tilde L_x \tilde L_y\,\eta^{\x} \, \eta^{\y} \Big]\,,\nn
A=&\ \frac{g'(r)^2}{4}a_t \eta^t+\frac{g'(r)}{g(r)}a_r dr+(\tilde L_x \,a_x -B \tilde L_x \tilde L_y\y )\eta^{\x}
+\tilde L_y a_y \eta^{\y}\,,\nonumber\\
\psi=&\ g(r)(\phi_1+i\phi_2)\,.
\end{align}
In these expressions we have used the one-forms  $\eta^t, \eta^{\x}, \eta^{\y}$ defined by
\begin{align}
\eta^t=dt+Q_{tr}dr+Q_{t\x} d\tilde x+Q_{t\y}d\tilde y\,,\quad \eta^{\x}=d\tilde x+Q_{r\x}dr\,,\quad \eta^{\y}=d\tilde y+Q_{r\y}dr\,.
%\eta^x=&dx+Q_{rx}dr\,,\nonumber\\
%\eta^y=&dy+Q_{ry}dr\,,\nonumber\\
\end{align}
The functions $f(r)$ and $g(r)$ are precisely the same functions appearing in the
Reissner-Nordstr\"om solution \eqref{eq:RN} and are incorporated for convenience. 
The remaining functions defined by
$\mathcal{F}\equiv \{Q_{tt},Q_{rr},Q,R,W,Q_{t \x},Q_{t\y},Q_{tr},Q_{r\x},Q_{r\y},a_t,a_r,a_{\x},a_{\y}, \phi_1,\phi_2\}$ are all functions of the radial coordinate $r$ as well as $(\tilde x,\tilde y)$, where the latter parametrise the torus with the identifications given in
\eqref{tildeidents}.
%\begin{align}
%f(r)&=\frac{(r-1)^2}{4 r_+^4}\left(B^2r^3(r-2)^3+r_+^2\left(\mu^2(r-2)^3 r^3+4(1+2r+3r^2-4 r^3+r^4)r_+^2\right)\right)\,,\nonumber\\
%g(r)&=\left(1-(1-r)^2\right)\,.
%\end{align}
On this torus, the scalar fields $\phi_1$ and $\phi_2$ satisfy the quasi-periodic boundary conditions \eqref{eq:bcscalars} while all the remaining functions in  $\mathcal{F}$ are periodic. There is significant redundancy in this ansatz and this will be dealt with 
momentarily.

We demand that the solutions approach an asymptotic $AdS_4$ boundary, located at $r = 0$. The boundary conditions
that we want to impose are given by
\begin{align}
&Q_{tt}(0,\x,\y)=Q_{rr}(0,\x,\y)=Q(0,\x,\y)=W(0,\x,\y)=1\,, \nn
&R(0,\x,\y)=v\,,\quad a_t(0,\x,\y)=\mu\,,\nonumber\\
&Q_{tr}(0,\x,\y)=Q_{r\x}(0,\x,\y)=Q_{r\y}(0,\x,\y)=Q_{t\x}(0,\x,\y)=Q_{t\y}(0,\x,\y)=0\,,\nonumber\\
&a_r(0,\x,\y)=a_{\x}(0,\x,\y)=a_{\y}(0,\x,\y)=\phi_1(0,\x,\y)=\phi_2(0,\x,\y)=0\,.
\end{align}
Notice that these boundary conditions imply that the asymptotic metric approaches \eqref{aasymmettwo}.
Furthermore, the asymptotic form of the gauge field is $A\to \mu dt -B\tilde L_x\tilde L_y \,\tilde{y}\,d\tilde{x}$, also as desired. Due to the presence of $g(r)$ in the
ansatz for $\psi$ in \eqref{ansatzpde}, the boundary conditions on $\phi_1$ and $\phi_2$ are associated with a
spontaneous breaking of the global $U(1)$ symmetry and we recall that we have assumed $\psi$ is dual to an operator with
$\Delta_\psi=2$. We will discuss the one point functions of this operator as well as the stress tensor and $U(1)$ current later.

We also demand that we have a Killing horizon, generated by the Killing vector $\partial_t$, located 
at $r=1$. This is achieved by demanding that the set of functions $\mathcal{F}(r,\x,\y)$, appearing
in \eqref{ansatzpde}, admit an expansion in powers of $(1-r)$ of the form
 \begin{align}
 \label{IRexp}
 \mathcal{F}=\mathcal{F}(1,\x,\y)-(1-r)\partial_r\mathcal{F}|_{r=1}+\dots\,.
 \end{align}
The equations of motion impose constraints on the coefficients appearing in \eqref{IRexp}. In particular, for the components $Q_{rr}(r,\x,\y)$ and $Q_{tt}(r,\x,\y)$
we obtain the condition for constant surface gravity $Q_{rr}(1,\x,\y)=Q_{tt}(1,\x,\y)$, while for all
other components we impose the Neumann boundary condition $\partial_r\mathcal{F}|_{r=1}=0$.
We also impose the following boundary condition on the complex scalar field at the horizon,
$\phi_2(1,1/2,1/2)=0$, which fixes the remaining Goldstone mode, as commented above.

At this point, we need to address the fundamental issue that the PDEs obtained after substituting the ansatz \eqref{ansatzpde}
into \eqref{eq:eom} are weakly elliptic, meaning that they are elliptic only for the physical degrees of freedom, and thus they are unsuitable for numerical solution without gauge fixing. To deal with this issue we use the DeTurck method, following \cite{Headrick:2009pv,Figueras:2011va}. We first modify the Einstein equations given in \eqref{eq:eom}, to obtain Einstein-DeTurck equations, by making the replacement 
 \begin{equation}\label{moddone}
 R_{\mu\nu}\to R_{\mu\nu}+\nabla_{\mu}\xi_{\nu}\,,\quad\text{with}\quad\xi^\mu=g^{\nu\lambda}(\Gamma^\mu_{\nu\lambda}(g)-\bar{\Gamma}^\mu_{\nu\lambda}(\bar{g}))\,.
 \end{equation}
Here 
%$\xi^\mu=g^{\nu\lambda}(\Gamma^\mu_{\nu\lambda}(g)-\bar{\Gamma}^\mu_{\nu\lambda}(\bar{g}))$ where 
$\bar{g}$ denotes a reference metric and $\bar{\Gamma}$ is the Christoffel connection of $\bar{g}$. We need to choose the reference metric to have a Killing horizon and the same asymptotic behaviour as the solutions we would like to construct; we will take it to be given by the metric in the dyonic AdS-RN black hole solution \eqref{eq:norm_ansatz}. For this choice, we find that $\xi^\mu=0$ on the boundary. After numerical integration we need to check, {\it  a posteriori}, that we have $\xi=0$ everywhere. The condition $\xi=0$ corresponds to fixing a set of coordinates. Similarly, we also need to fix the gauge freedom for the gauge-field.
As in \cite{Donos:2015eew} (see also \cite{Withers:2014sja}) we modify the Maxwell equation in \eqref{eq:eom} via
\begin{align}\label{moddtwo}
\nabla_{\mu}F^{\mu\nu}\to \nabla_{\mu}F^{\mu\nu}+ \nabla^\nu\varphi\,,\quad\text{with}\quad
 \varphi=\nabla_\mu A^\mu +\xi_\mu A^\mu - g^{\mu\nu}\bar{\nabla}_\mu\bar{A}_\nu\,,
\end{align}
where $\bar{A}_\nu$ is a reference gauge field. 
While $\bar{A}_\nu$ is not needed in order to obtain an elliptic system
of PDEs, it does allow us to suitably choose $A_\mu-\bar A_\mu$ for our boundary value problem. 
Indeed we take $\bar{A}=-B \tilde L_x\,\tilde L_y \y d\x$ so
that $A-\bar A$ are periodic functions and, as one can then explicitly check, $\varphi$ then 
vanishes on the boundary. As in \cite{Donos:2015eew}, by
taking the divergence of the modified Maxwell equation and then integrating over the bulk, one can then deduce
that $\varphi=0$ everywhere. The vanishing of $\varphi$ fixes the gauge invariance.

To summarise, to obtain our vortex lattice solutions, we will solve the modified equations of motion, given in 
\eqref{moddone},\eqref{moddtwo}, which gives a system of elliptic PDEs with a well defined boundary value problem. Having solved
them we then check that $\xi^\mu=0$ and since $\xi^\mu$ is spacelike, this is achieved by checking $\xi^2=0$.
Some additional comments on the numerical approach we take, as well as a discussion of the numerical convergence is given in appendix \ref{appendixB}.

Returning to our ansatz, \eqref{ansatzpde}, we see that it is left invariant if we make the replacements:
$\tilde L_x\leftrightarrow \tilde L_y$, $W\to W^{-1}$,  $Q_{r\tilde x} \leftrightarrow Q_{r\tilde y}$, $Q_{t\tilde x}  \leftrightarrow Q_{t\tilde y} $,
$a_{\tilde x}\to a_{\tilde y}+B\tilde L_y \tilde y$, $a_{\tilde y} \to a_{\tilde x}-B \tilde L_x \tilde y$ and interchange $\tilde x\leftrightarrow \tilde y$. This symmetry is then reflected in our solutions. In particular, after recalling \eqref{kdef}, it implies that any physical quantities that
are obtained by integrating over $\tilde x, \tilde y$, such as the free energy, for example, will be invariant
under the interchange of $k\to 1/k$.

\subsection{Thermodynamics and one point functions}
The thermodynamic properties of the black holes are obtained by analytically continuing the time
coordinate via $t=-i\tau$. Demanding regularity of the solution at the black hole horizon we obtain the temperature, $T$, which has the same form as given in \eqref{tempexp}.  We can also read off the area of the event horizon and, since we are working in units with $ 16 \pi G=1$, we deduce that the entropy density is given by the horizon integral
\begin{equation}
s=4 \pi\,r_+^2\,\int d\x d\y Q|_{r=1}\,.
\end{equation}
where the integral is over a unit cell.

To calculate the free energy we need to consider the total Euclidean action, 
$I_{Tot}=I+I_{bdr}$, where $I=-iS$, with $S$ as in \eqref{eq:action},  and $I_{bdr}$  
given by the following integral on the boundary $r\to 0$:
\begin{equation}\label{ctermp}
I_{bdy}=\int d\tau d\x d\y\sqrt{\gamma}(-2 K+4+\dots)\,.
\end{equation}
Here $K$ is the trace of the extrinsic curvature of the boundary and $\gamma_{\mu\nu}$ is the 
induced boundary metric 
given in \eqref{aasymmettwo} up to a factor of $r_+^2/(4r^2)$ and the dots refer to a term quadratic in the scalar field which does not play a role for the solutions
that we consider in this paper.
To obtain the free energy density, $w$, we write the total free energy as $T[I_{Tot}]_{OS}\equiv w vol_2$.

It is similarly straightforward to obtain the expectation values for the boundary stress tensor, $T^{mn}$, and the abelian current vector,
$J^m$, as well as the condensate $\langle \mathcal{O}_\psi\rangle$ (we have presented a few details in
appendix \ref{appendixA}). Since the solutions have a timelike Killing vector $\partial_t$, we have
\begin{align}
\label{eq:OSactionp}
w&= \bar T^{tt}-s T-\mu  \bar J^t\,,
\end{align}
where the bar refers to a period average over the spatial coordinates e.g. 
\begin{align}
\bar J^t=\int d\x d\y\sqrt{\gamma}J^t\,.
\end{align}

The free energy density $w$ depends on $T,\mu,B$, the integer $n$ (giving the number of vortices per unit cell), and also on the shape of the lattice, which can be specified by
$v,k$, where we recall $k$ was defined in \eqref{kdef}: $w=w(T,\mu,B;v,k;n)$. 
To see how the free energy depends on varying $B,v$ and $k$ we follow 
\cite{Donos:2013cka}. A calculation\footnote{To see this, it is convenient to work in the $\tilde x,\tilde y$ coordinates and consider the free parameters to be $\tilde L_x$, $\tilde L_y$ and $v$. As we vary these parameters we vary the boundary metric \eqref{eq:bmetric_num} 
and we also notice that the boundary gauge field in \eqref{eq:bmetric_num2} is given by
$A_{\partial}=\mu dt-\frac{2\pi n}{q}\,\tilde y\,d\tilde x$ and hence is independent of these parameters. From (2.13) of
\cite{Donos:2013cka} we deduce
\begin{align}
\delta w&=-\frac{\delta\tilde L_x}{\tilde L_x}(w+\bar T^{\x}{}_{\x})
-\frac{\delta\tilde L_y}{\tilde L_y}(w+\bar T^{\y}{}_{\y})
+\frac{\delta v}{(1+v^2)}\Big(\frac{v}{2}(\bar T^{\x}{}_{\x}-\bar T^{\y}{}_{\y})-\frac{\tilde L_x}{\tilde L_y}\sqrt{1+v^2}\bar T^{\x}{}_{ \y}\Big)\,,
%\nn
%&=\frac{\delta\tilde L_x}{\tilde L_x}(w+\bar T^{\x}{}_{\x})
%+\frac{\delta\tilde L_y}{\tilde L_y}(w+\bar T^{\y}{}_{\y})
%-\frac{\delta v}{(1+v^2)}\Big(\frac{v}{2}(\bar T^{\y}{}_{\y}-\bar T^{\x}{}_{\x})-\frac{\tilde L_y}{\tilde L_x}\sqrt{1+v^2}\bar T^{\y}{}_{ \x}\Big)\,,\nonumber
\end{align}
and we then use \eqref{kdef}.
} shows that we can write 
\begin{align}\label{varyvlat}
\delta w=
\frac{\delta k}{k}(\bar T^{\x}{}_{\x}-\bar T^{\y}{}_{\y})
&+\frac{\delta B}{2B}(2w+\bar T^{\x}{}_{\x}+\bar T^{\y}{}_{\y})\nn
&+\frac{\delta v}{(1+v^2)}\Big(\frac{v}{2}(\bar T^{\x}{}_{\x}-\bar T^{\y}{}_{\y})-\frac{\tilde L_x}{\tilde L_y}\sqrt{1+v^2}\bar T^{\x}{}_{ \y}\Big)\,.
\end{align}

In order to find the thermodynamically preferred black holes for given UV data $(T,\mu,B)$ and given $n$, we need to minimise
$w$ over the lattice parameters $(v,k)$. 
From \eqref{varyvlat} the preferred vortex lattice will therefore satisfy 
$\bar T^{\x}{}_{\x}=\bar T^{\y}{}_{\y}$, $T^{\x}{}_{ \y}=0$. Since the symmetry of the stress tensor implies that
$\sqrt{1+v^2}(\frac{\tilde L_y}{\tilde L_x}\bar T^{\y}{}_{ \x} -\frac{\tilde L_x}{\tilde L_y}\bar T^{\x}{}_{ \y})={v}(\bar T^{\y}{}_{\y}-\bar T^{\x}{}_{\x})$, we also have 
$T^{\x}{}_{ \y}=0$. Thus, we deduce that
\begin{align}\label{gresone}
\bar T^i{}_j=p\delta^i{}_j\,,\qquad p=-\frac{1}{2}\bar T^t{}_t\,,
\end{align}
where we also used tracelessness of the stress tensor.
As in \cite{Donos:2013cka} we can also conclude that
\begin{align}\label{grestwo}
\bar Q^{i}&\equiv -(\bar T^i{}_t+\bar J^{i}\mu)=0\,,\nn
\bar J^{i}&=0
\end{align}
These conditions are a result of the fact that in thermal equilibrium
the spatial parts of the heat current, $Q^i$, and the abelian current, $J^i$, are
magnetisation currents of the form $\sqrt{\gamma}J^i=\partial_j M^{ij}$ and
$\sqrt{\gamma}Q^i=\partial_j M_T^{ij}$ where 
$M^{ij}=M^{[ij]}$ and $M_T^{ij}=M_T^{[ij]}$ is the local magnetisation density and the
thermal magnetisation density, respectively. 
A subtlety is that in a superfluid state we can have $\bar J^i\ne 0$ but with $\bar Q^i=0$, but
for the thermodynamically preferred configurations we have $\bar J^{i}=0$ \cite{Donos:2018kkm}.

Finally, we note that using the above results, the first law for the thermodynamically preferred vortex
lattice black holes can be written as
\begin{align}\label{firstlaw}
\delta w=&-s \delta T-\bar J^t\delta\mu -m \delta B\,,
\end{align}
where 
\begin{align}\label{magexp}
&m=-\frac{w+\bar p}{B}\,.
\end{align}
In our numerical results, to be discussed next, we have directly verified that \eqref{gresone}-\eqref{magexp}
all hold.

\section{Numerical Results}\label{sec:Results}

In the solutions we have numerically constructed which we discuss below, we have fixed\footnote{We also carried out a check of \eqref{firstlaw}, \eqref{magexp} by constructing some black holes with different values of $B$.} 
\begin{align}\label{beeval}
q=2,\qquad B/\mu^2=0.01\,.
\end{align}
We express physical quantities in
terms of the chemical potential $\mu$; this can be achieved by setting $\mu=1$ and then reinstating $\mu$ afterwards using dimensional
analysis. For this value of $|qB/\mu^2|$ the critical temperature at which the dyonic AdS-RN black hole become unstable 
to zero modes associated with the lowest Landau level is  $T_c/\mu=0.088$. We constructed the new branch of vortex lattice black 
holes that exist below $T_c$ which are parametrised by three continuous parameters, the temperature, $T$, and two parameters, $v,k$, 
which determine the shape of the lattice  (recall $k$ was defined in \eqref{kdef}). 
We also note that the plots of the free energy that we present below are invariant under $k\to 1/k$
for reasons discussed at the end of section \ref{sec:numerics}. 
In principle, the solutions are also specified by a discrete parameter $n$ that appears in the flux quantisation condition
\eqref{qconag},
but all of the solutions we discuss below\footnote{We looked for solutions with $n\ne 1$, but just found 
rescaled versions of solutions with $n=1$.} have $n=1$.

\begin{figure}[h!]
\centering
\includegraphics[width=0.45\linewidth]{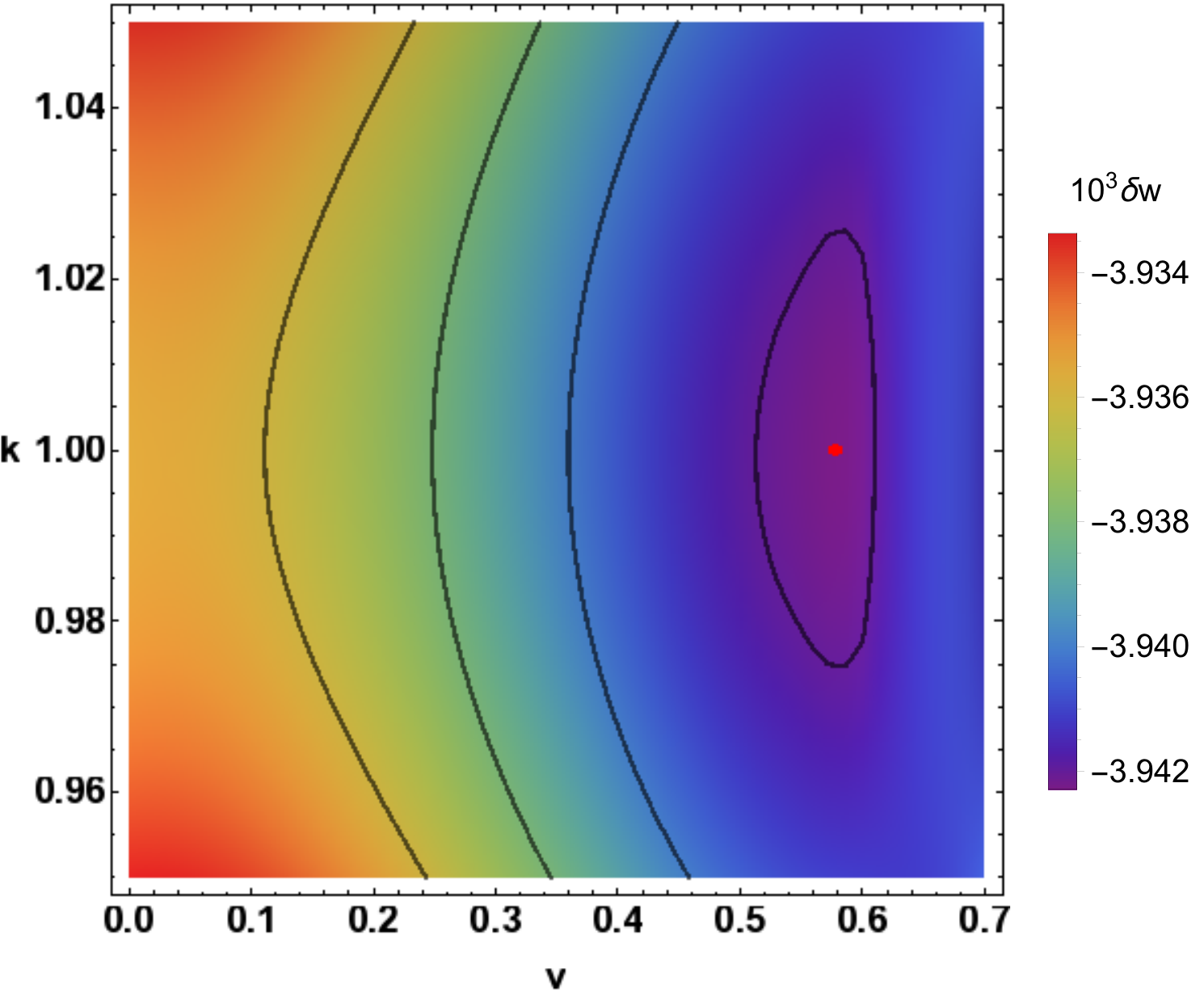}
\includegraphics[width=0.45\linewidth]{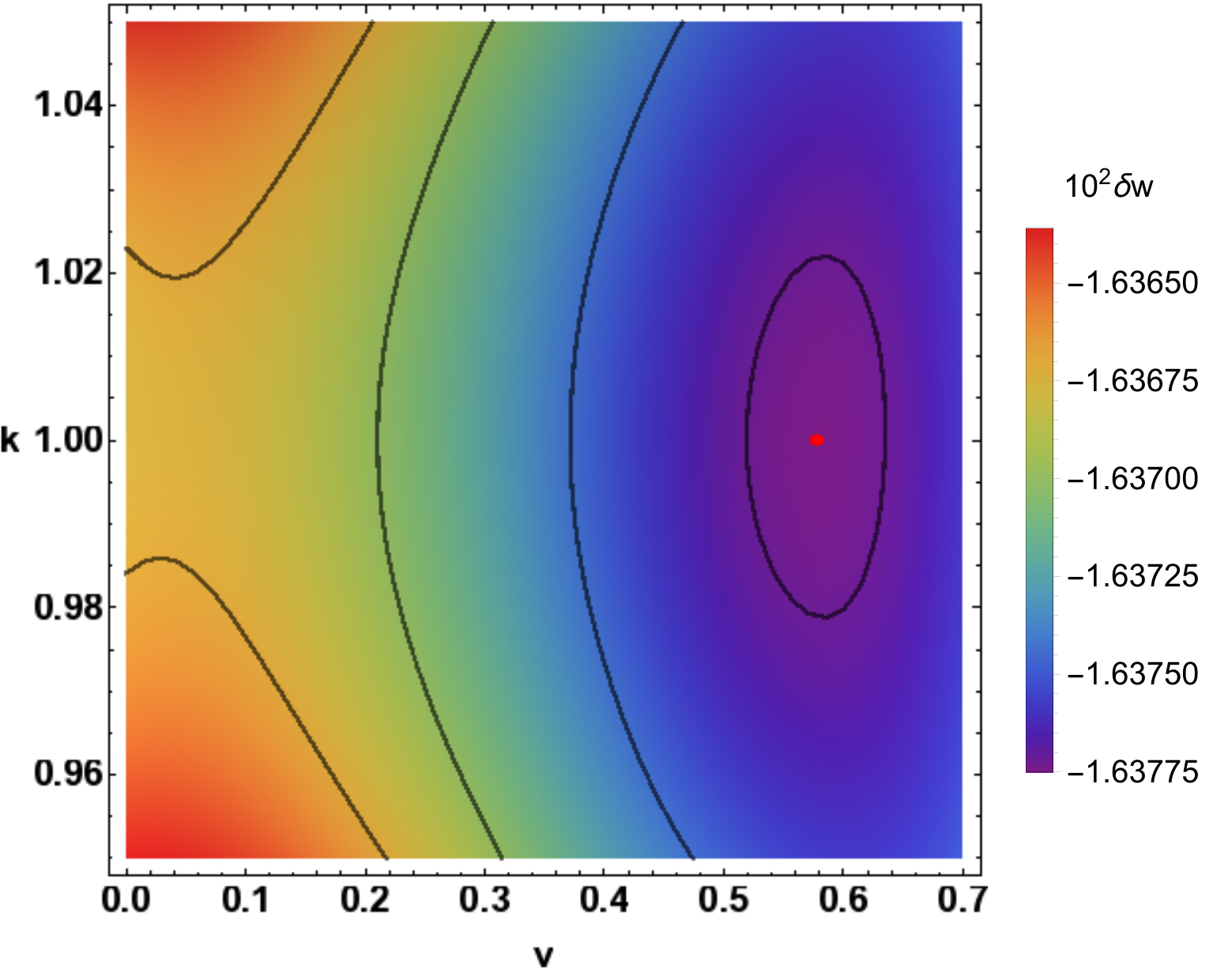}
\caption{Plots displaying the behaviour of the free energy density for the vortex lattice black holes
(lowest Landau level) as a function of the shape, parametrised by $(v,k)$, for
 with $T/T_c\sim0.79$ and $T/T_c\sim0.565$. We have plotted $\delta w/\mu^3$, the difference between
the free energy of the lattice and of the normal phase (the AdS-RN solution).
The red dot corresponds to the configuration minimising the free energy and it describes a triangular vortex lattice with $v=1/\sqrt{3}$ and $k=1$ (equivalently $\tilde{L}_x=\tilde{L}_y$). }
\label{fig:FE}
\end{figure}

We have constructed several black hole solutions, associated with the lowest Landau level, for various values of the parameters $T,v,k$, but we studied in more depth three
specific values of the temperature: $T/\mu=0.08$, $T/\mu=0.07$ and $T/\mu=0.05$ with $T/T_c\sim0.905, 0.79$ and $0.565$, respectively. 
For each of these temperatures we determine the shape of the thermodynamically preferred configuration by calculating the free energy density $w/\mu^3$ as a function of $(v,k)$. For all these temperatures, within numerical precision, we find that the preferred black holes have $k=1$ and $v=1/\sqrt{3}$, corresponding to a triangular vortex lattice. This is clearly illustrated in figure \ref{fig:FE} for $T/T_c\sim0.79$ and $T/T_c\sim0.565$. 

\begin{figure}[h!t]
 \textbf{\hspace{2.5cm} $ \langle |\mathcal{O}_{\psi}|\rangle/\mu^2$\hspace{6cm} ${J^i/\mu^2}$}\par\medskip
\includegraphics[width=0.48\linewidth]{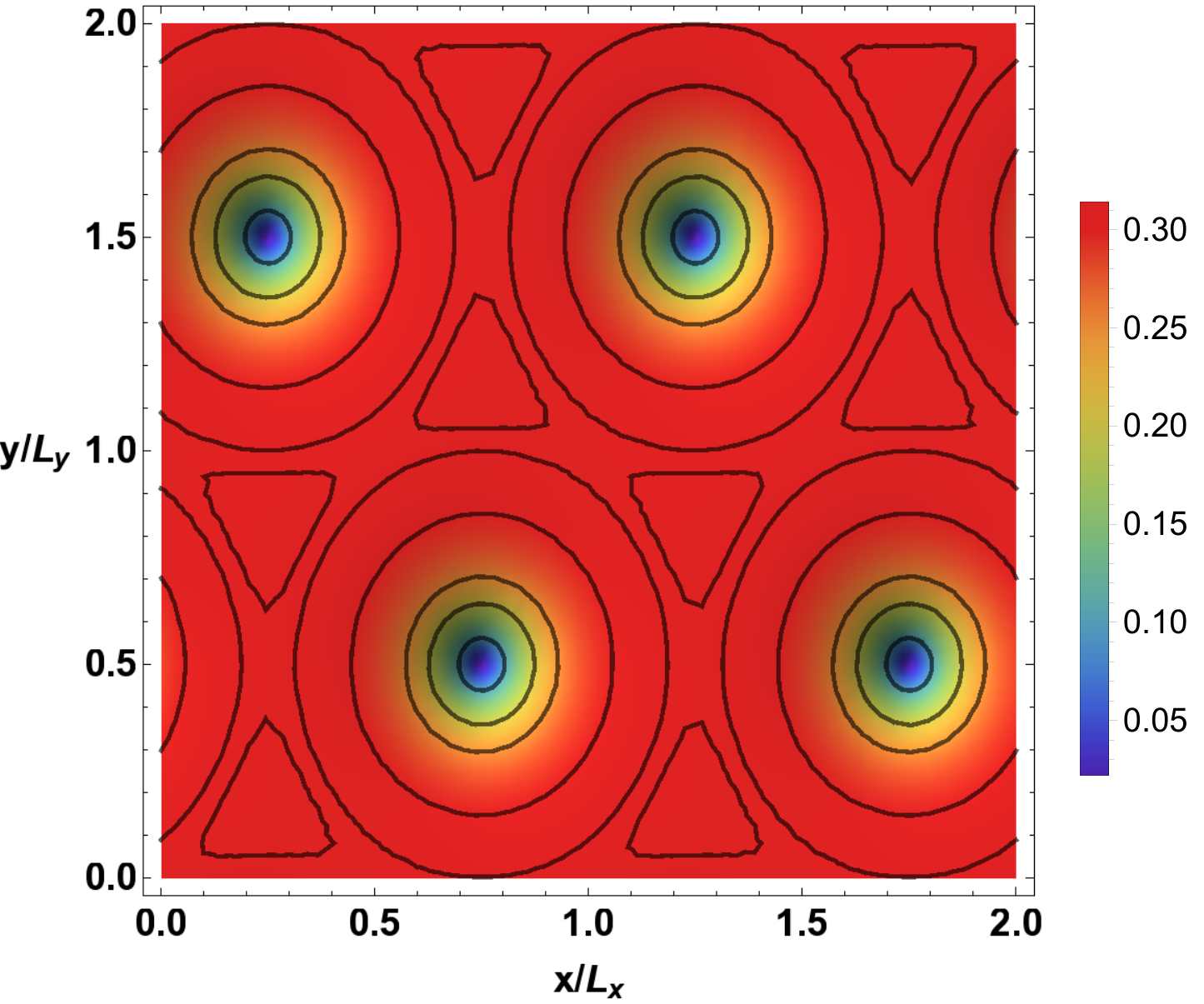}\quad \includegraphics[width=0.48\linewidth]{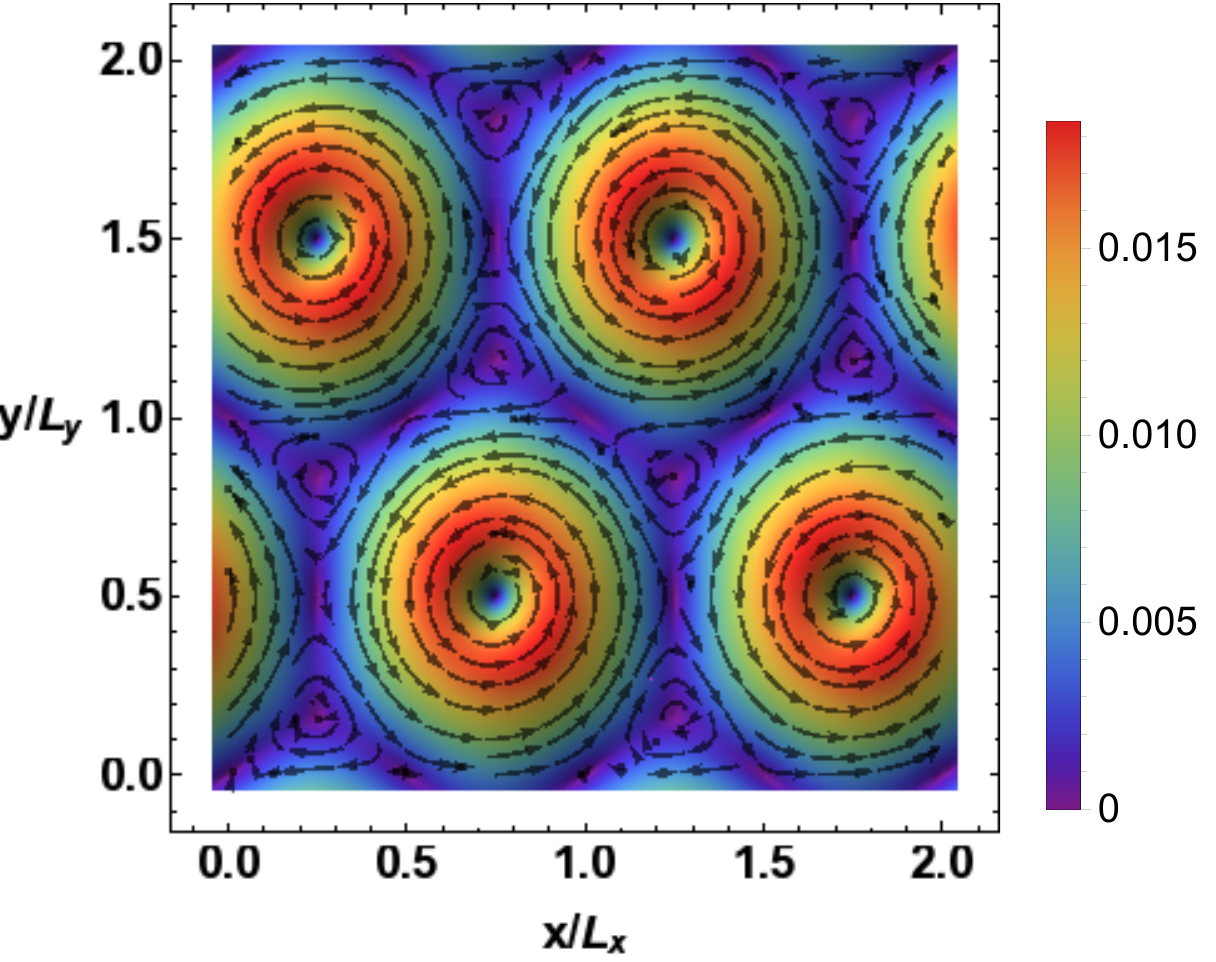}\par\medskip
 \textbf{\hspace{2.5cm} $ T^{tt}/\mu^3$\hspace{6.5cm} $J^t/\mu^2$}\par\medskip
\includegraphics[width=0.48\linewidth]{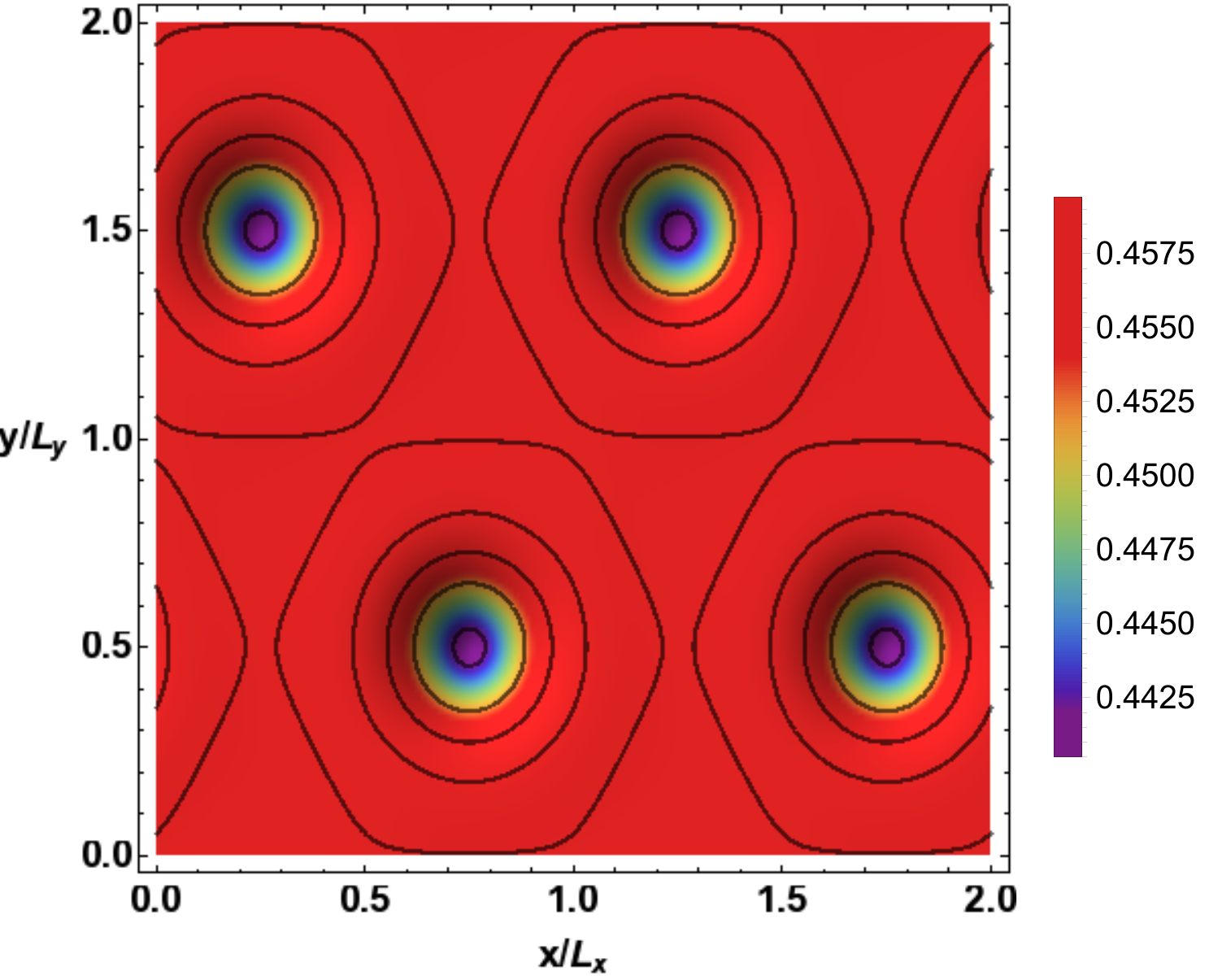}\quad \includegraphics[width=0.48\linewidth]{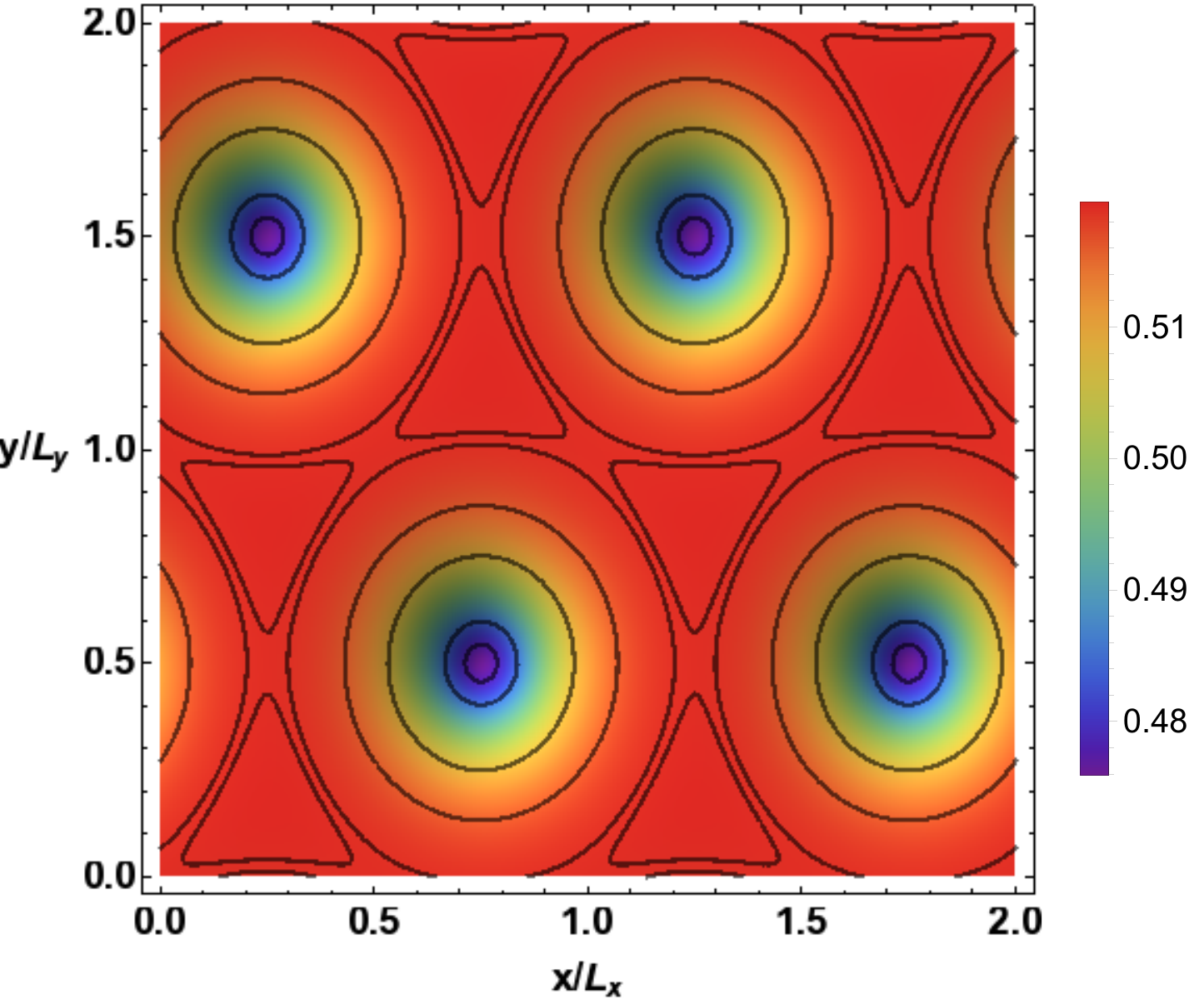}
\caption{Observables for the preferred triangular vortex lattice black holes (lowest Landau level)
for $T/T_c\sim0.79$. 
Here $\langle |\mathcal{O}_{\psi}|\rangle$ is the order parameter, $J^i$ is the local current density, 
$T^{tt}$ is the energy density and $J^t$ is the charge density.
Note that these quantities are plotted in the original, untransformed spatial coordinates $(x,y)$,
scaled by $L_x$ and $L_y$ (see \eqref{eq:torus} and \eqref{tilcchge}).
}
\label{fig:observables}
\end{figure}

We can also calculate various physical observables in the dual field theory for the preferred triangular vortex lattice black holes. 
In figure \ref{fig:observables} we present the plots for the case $T/T_c\sim0.79$. We see that the 
modulus of the order parameter, $\langle |\mathcal{O}_{\psi}|\rangle$, has zeroes at the centre of each vortex, as expected. Furthermore, we have explicitly checked that
the phase of the order parameter winds exactly once around each vortex i.e. $n=1$.
The spatial components of the local current density $J^i$
are magnetisation currents (i.e. with no net transport, $\bar J^i=0$) and circulate around each vortex core. Since we are considering a 
superfluid these currents do not diminish the background homogeneous magnetic field, which is the constant value of $B$ throughout. 
The charge density $J^t$ is non-vanishing at the core of the vortices 
and is associated with the non-vanishing charge density of
the normal phase. The plots also show that the 
energy density $T^{tt}$ and  $J^t$ are slightly 
diminished\footnote{From a Landau-Ginzburg description, which should be valid near
the critical temperature, one expects
the charge density will be reduced at the cores. This is because the charge density is given by the charge density of the normal 
phase plus a contribution from the norm squared of the order parameter. }at the cores of the vortices.

\begin{figure}[h!]
 \textbf{\hspace{2.7cm} $ |\psi_h|$}\par\medskip
\includegraphics[width=0.45\linewidth]{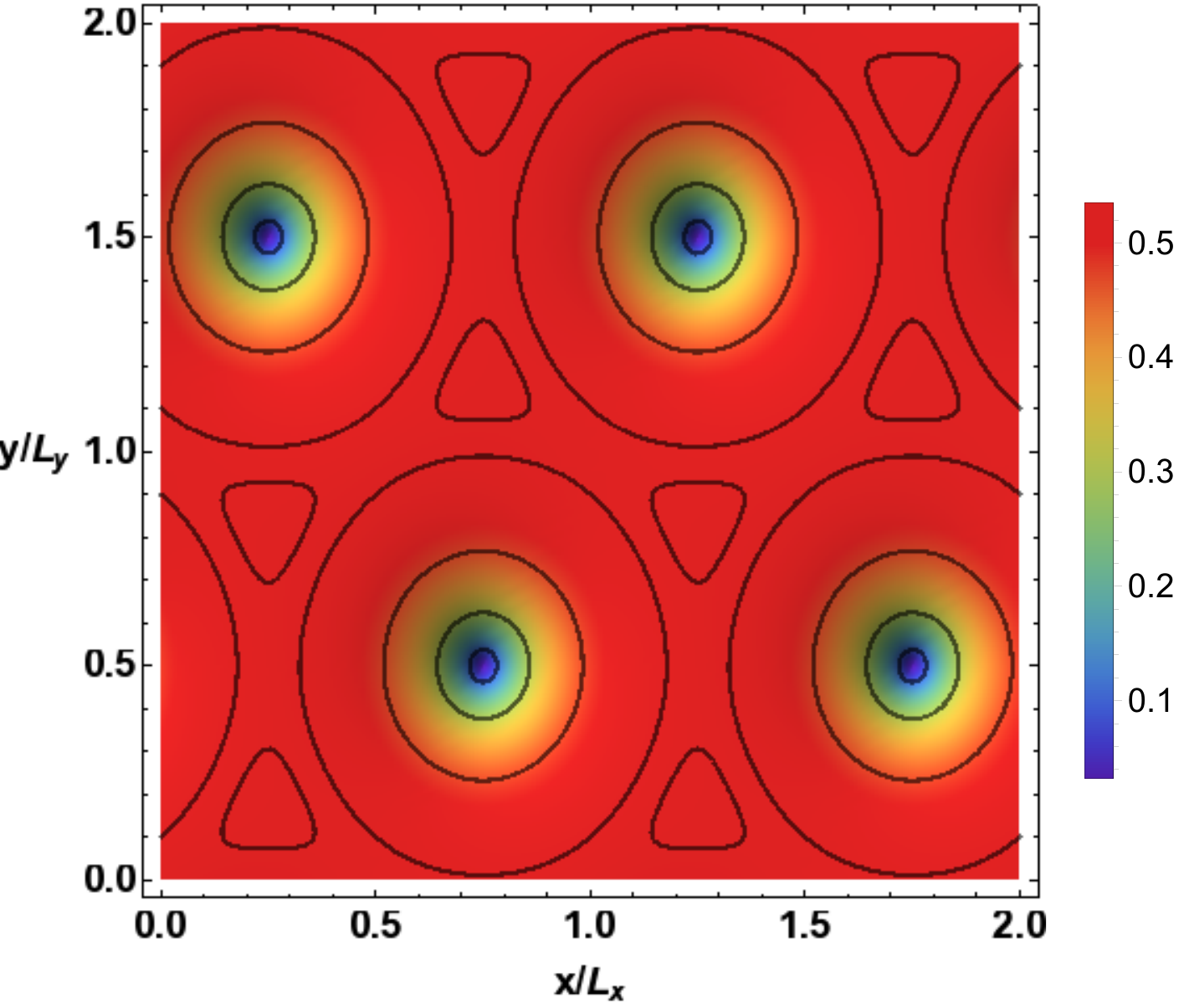}\quad
\includegraphics[width=0.48\linewidth]{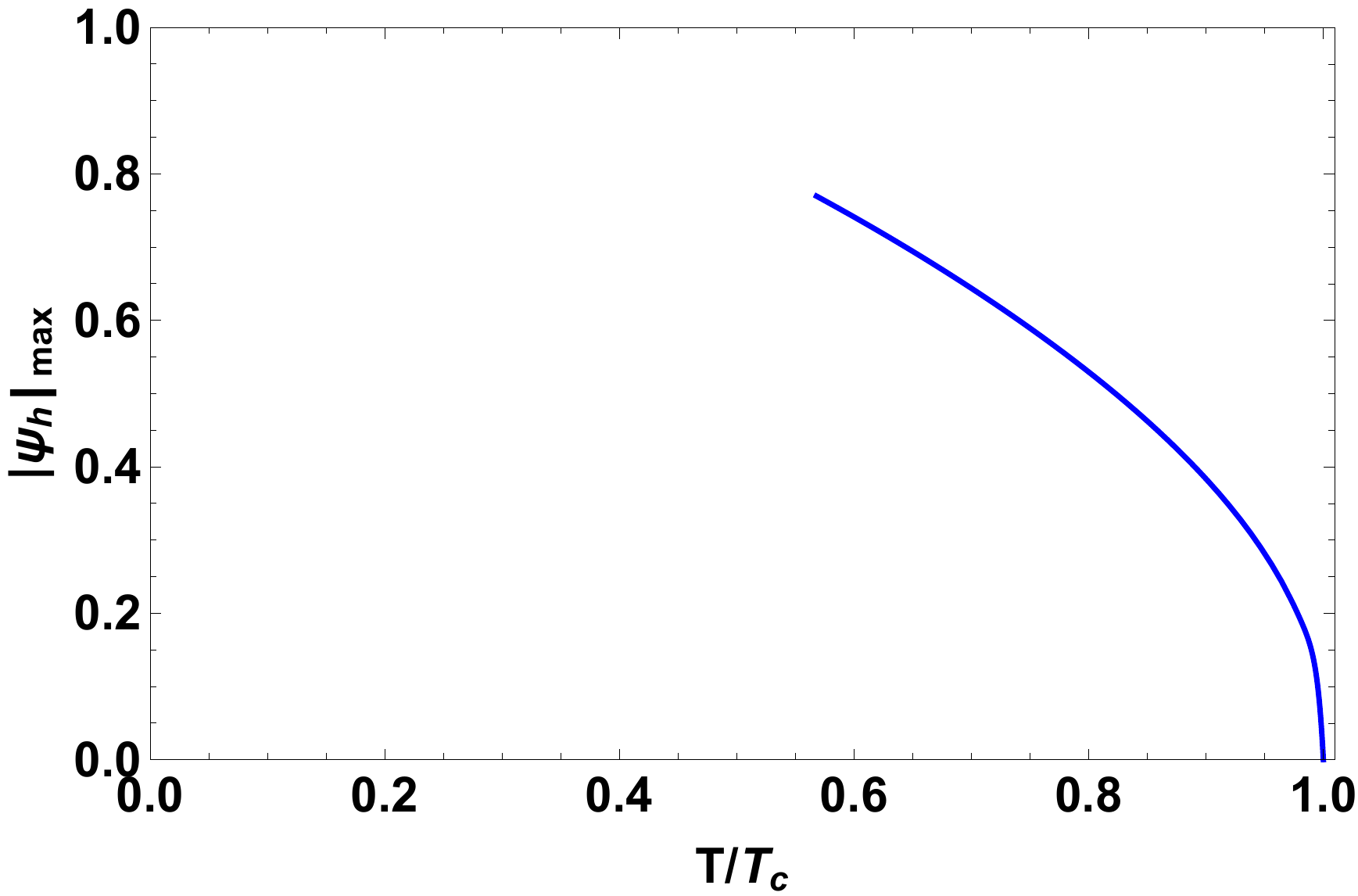}
\caption{(a)Plot of the modulus of the scalar field $|\psi_h|$ at the horizon for $T/T_c=0.79$ 
for the preferred triangular vortex lattice black holes (lowest Landau level). (b) Plot of maximum value of the modulus of the scalar field, 
$|\psi_h|_\text{max}$, against $T/T_c$ for the same black holes.}
\label{fig:psi}
\end{figure}

It is also interesting to examine the structure of the preferred black hole solutions themselves. In particular, the black hole horizons have an inhomogeneous geometry, with peaks that are associated with the position of the vortices.
A complementary picture can be illustrated by studying the modulus of the scalar field, $|\psi_h|$, on the horizon, as illustrated
in figure \ref{fig:psi}(a). At the core of the peaks, the value of $|\psi_h|$ is zero.
Furthermore, the maximum value, $|\psi_h|_\text{max}$, which is reached at the midpoint between
two peaks monotonically increases as the temperature is lowered, as shown in figure \ref{fig:psi}(b), 
leading to more pronounced peaks. 

Although we are some way from zero temperature, the behaviour in figure \ref{fig:psi}(b) suggests that 
$|\psi_h|_\text{max}$ is approaching one. Furthermore, we find that the temperature dependence of, for example, the 
maximal value of the norm of the gradient of the scalar field as well as $F_{\mu\nu}F^{\mu\nu}$
at the horizon are consistent with them staying finite as $T\to 0$.
This indicates that at zero temperature there will be a regular\footnote{This should be contrasted with what seems to be happening with the solutions discussed in \cite{Donos:2015eew}. We also note that constructions
of extremal black holes with inhomogeneous horizons, parametrised by spatially periodic scalar fields, have been discussed in \cite{Donos:2013eha,Blake:2016jnn,Donos:2017sba}.
}, extremal horizon
with an inhomogneous peaked structure with $|\psi|$ interpolating between
$|\psi|=0$ at the core of the vortices to a maximum value of $|\psi|=1$, the value associated with the symmetry breaking $AdS_4$ solution,
between the vortices.
It would be very interesting to explicitly construct these putative $T=0$ solutions.

\begin{figure}[h!]
\centering
\includegraphics[width=0.45\linewidth]{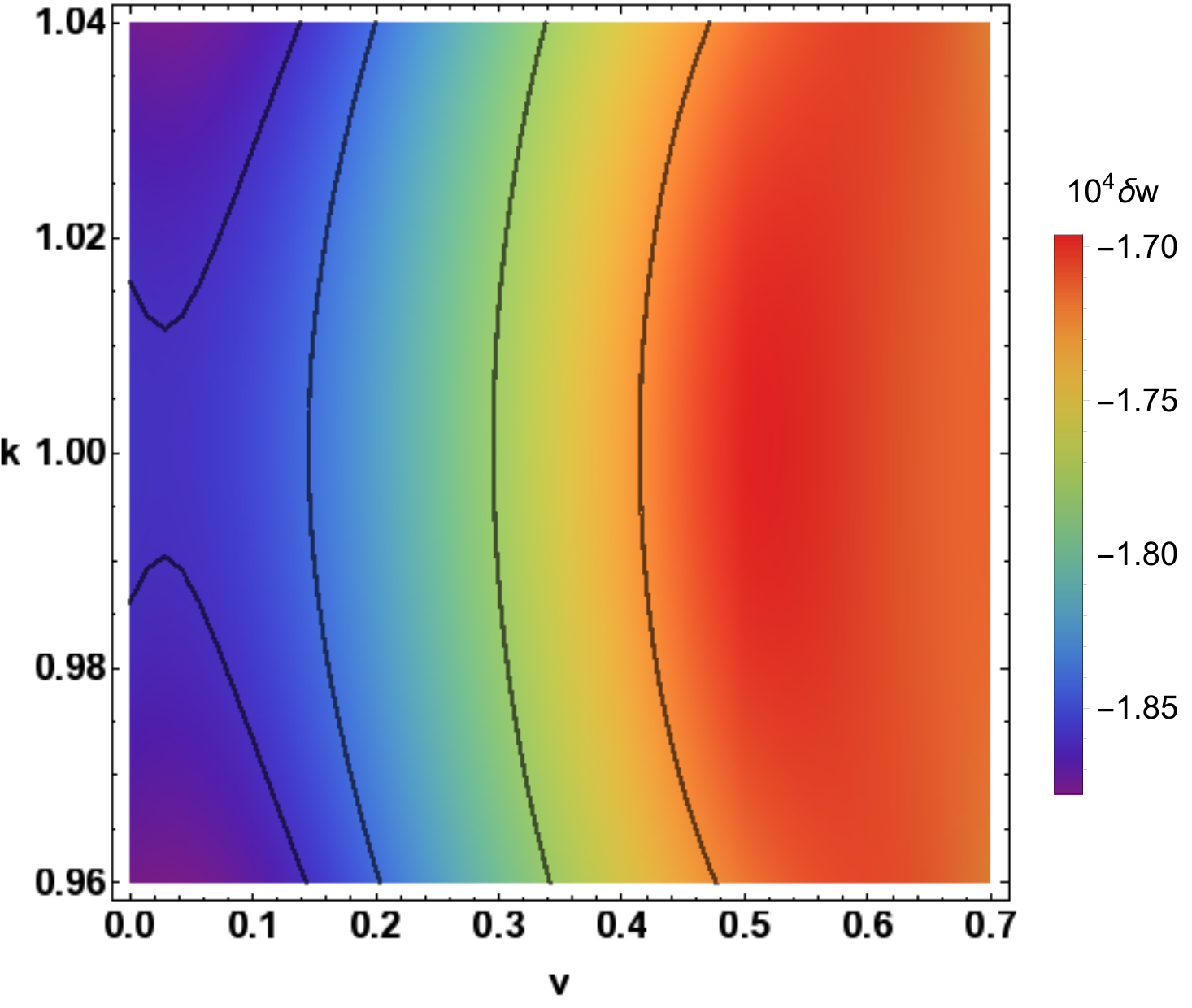}\quad\includegraphics[width=0.48\linewidth]{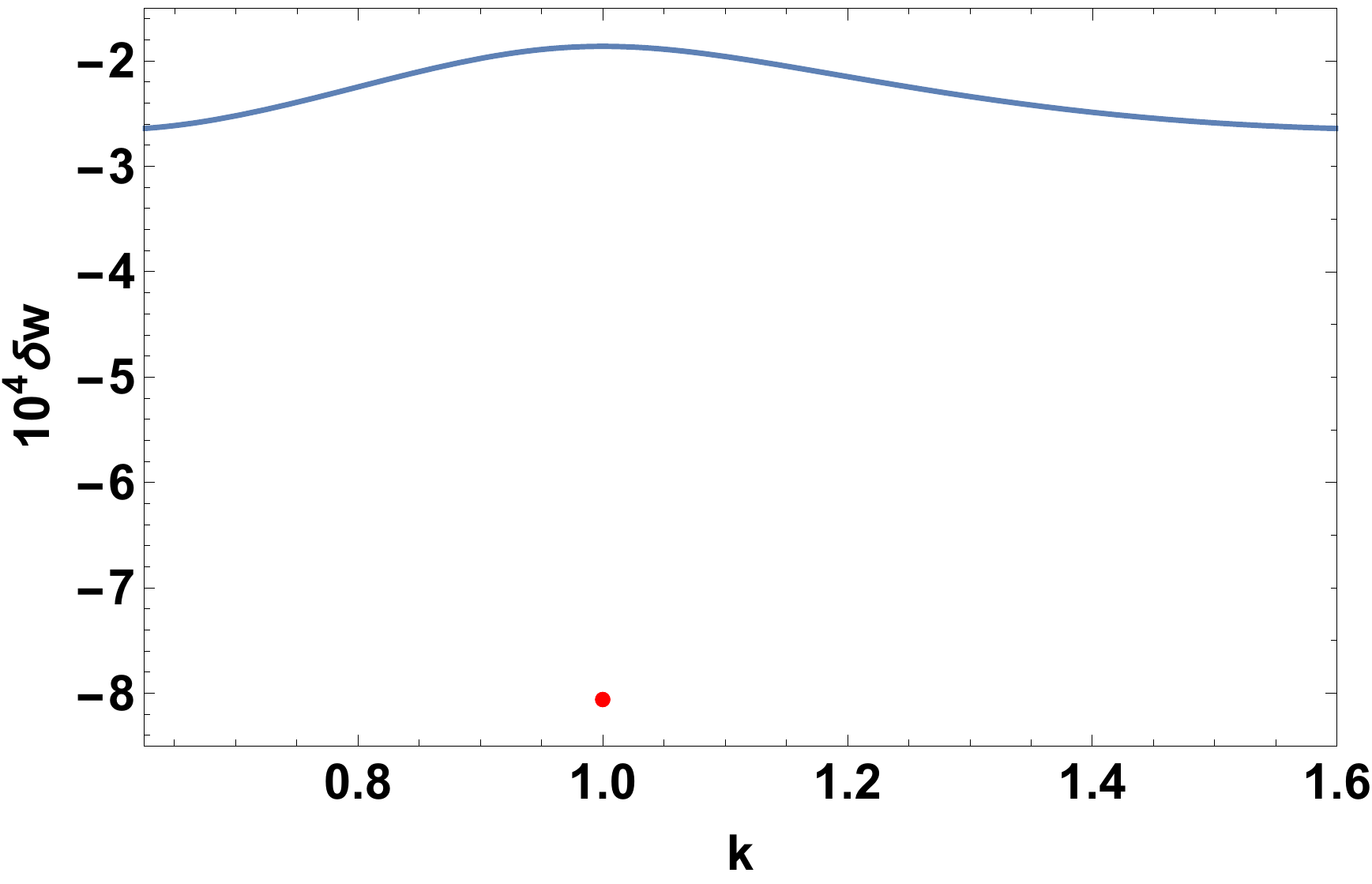}
\caption{Plots displaying the behaviour of the free energy density 
for vortex lattice black holes associated with the first Landau level and temperature $T/\mu=0.08$.
We have plotted $\delta w/\mu^3$, the difference between the free energy of the lattice and of the normal phase (the AdS-RN solution).
(a) Plot of $\delta w/\mu^3$ as a function of $k$ and $v$. 
(b) Plot of $\delta w/\mu^3$ as a function of $k$ for $v=0$, with the red dot indicating the value of the free energy
for the thermodynamically preferred triangular vortex lattice associated with the lowest Landau level.
The two plots indicate that the minimum of the free energy for the first Landau level
is when $v=0$ and $k\to 0$ or $k\to \infty$, corresponding to the unit cell becoming infinitely long and thin.}
\label{fig:FLL}
\end{figure}

For $T/\mu\sim 0.085$ the original dyonic AdS-RN black hole solution becomes unstable to zero modes
associated with the first Landau level (i.e. $j=1$ in figure \ref{fig:Tc}). Thus below this temperature, there is an additional family of 
vortex lattice black hole solutions, again parametrised by the temperature $T$ and two shape parameters $v,k$
(and $B$ fixed as in \eqref{beeval}). 
We have constructed some examples of these back-reacted black holes for $T/\mu=0.08$ and various values of $k$ and $v$ and we find, surprisingly, that the preferred black holes in this branch appear to have $v=0$ and $k\to 0$ or $k\to \infty$, corresponding to the unit cell becoming infinitely long and thin (see figure \ref{fig:FLL}). This suggests that the vortex lattice might be trying to form a linear defect. However, we also find that the black holes associated with the first Landau level that we have constructed are never thermodynamically preferred
when compared with any of the black holes that are associated to the lowest Landau level that we discussed above and, in particular, the preferred triangular vortex lattice. 

\section{Discussion}\label{disc}
For a specific holographic model we have constructed fully back-reacted black holes that are dual to a lattice of vortices 
in a superfluid phase in the presence of a homogeneous, constant magnetic field. We have shown that the thermodynamically preferred
black holes describe a triangular lattice of vortices and associated with the lowest Landau level,
at least for the temperatures we have considered. 

Assuming that this picture persists to lower temperatures, which seems most likely to us, it appears that 
at zero temperature there are novel, non-singular extremal black holes, which are not static, with
the core of each vortex on the horizon approaching a dyonic $AdS_2\times\mathbb{R}^2$ configuration and these are embedded in an ambient sea of the IR $AdS_4$ superfluid
ground state, in the presence of a uniform magnetic field. It would be of much interest to construct these extremal black holes directly.
As a first step, it could
be worthwhile to first try and construct a kind of inside-out near horizon solution which describes a bubble of $AdS_4$ in the superfluid ground state, embedded in an ambient dyonic $AdS_2\times\mathbb{R}^2$ solution\footnote{This would be analogous to the interesting defect
solutions
constructed in \cite{Dias:2013bwa}. There a holographic setup
with vanishing background magnetic field and zero chemical potential was considered, with the superconducting instability induced via a double trace deformation. The absence of 
a background magnetic field  meant that it was possible to construct a single vortex 
solution with a bubble of magnetic $AdS_2\times\mathbb{R}^2$ embedded in an
$AdS_4$ symmetry breaking vacuum}.

There are many natural ways to modify the constructions reported in this paper. While we suspect that we will get similar results by simply varying $q$
and $B$, this may not be the case and other shapes of vortex lattices could in fact be preferred. 
We think it would be particularly interesting to extend the model to include a neutral scalar $a$ with an $aF\wedge F$ coupling in the bulk Lagrangian. On the one hand such a term arises in the top down models considered in \cite{Gauntlett:2009bh,Donos:2012yu}
which also explored a fascinating competition between superfluid phases and phases in which the 
$U(1)$ symmetry is unbroken but spatial translations are spontaneously broken.
In addition given the fact that in field theory a Chern Simons term gives rise to novel vortices carrying both electric and magnetic charges (e.g. see \cite{Horvathy:2008hd}), which are models for anyons, we can expect that the associated gravitational construction\footnote{A calculation in the probe approximation for a single vortex was carried out in \cite{Tallarita:2010vu}.} may well exhibit novel phenomena too. 

The thermodynamically preferred black holes that we constructed in this paper are associated with vortices in
the lowest Landau level. However, this is not expected to be the case for other models such as the ones considered in
\cite{Almheiri:2014cka,Donos:2011pn}. These black holes could have a very interesting structure based on the fact 
that the preferred black hole vortex lattice in the first Landau level which we constructed here has an interesting linear defect structure.

\subsection*{Acknowledgments}
This work made use of the facilities of the Hamilton HPC Service of Durham University.   AD is supported by STFC grant ST/P000371/1. JPG is supported by the European Research Council under the European Union's Seventh Framework Programme (FP7/2007-2013), ERC Grant agreement ADG 339140. JPG is also supported by STFC grant ST/P000762/1, EPSRC grant EP/K034456/1, as a KIAS Scholar and as a Visiting Fellow at the Perimeter Institute. JPG thanks the KITP, UCSB for hospitality:  this research was supported in part by the National Science Foundation under Grant No. NSF PHY-1748958.
C.P. is supported by the European Union's Horizon 2020 research and innovation programme under the 
Marie Sk\l odowska-Curie grant agreement HoloLif No 838644.

\appendix

\section{Asymptotic Expansion}\label{appendixA}
Using the modified equations of motion as described in section \ref{sec:numerics}, the asymptotic boundary expansion 
for the functions appearing in the ansatz \eqref{ansatzpde} take the form
\begin{align}\label{eq:UVexp}
&Q_{tt}(r,\x,\y)=1+r^3c_{tt}(\x,\y)+g_1(\x,\y) r^{(3+\sqrt{33})/2}+\mathcal{O}(r^4)\,,\nonumber\\
&Q_{rr}(r,\x,\y)=1+g_2(\x,\y) r^{(3+\sqrt{33})/2}+\mathcal{O}(r^4)\,,\nonumber\\
&Q(r,\x,\y)=1-\frac{1}{2}r^3c_{tt}(\x,\y)+g_1(\x,\y) r^{(3+\sqrt{33})/2}+\mathcal{O}(r^4)\,,\nonumber\\
&W(r,\x,\y)=1+r^3 c_W(\x,\y)+\mathcal{O}(r^4)\,,\quad R(r,\x,\y)=v+r^3c_R(\x,\y)+\mathcal{O}(r^4)\,,\nonumber\\
&Q_{tr}(r,\x,\y)=r^4 c_{tr}(\x,\y)+\mathcal{O}(r^4 \,lnr)\,,\quad Q_{t\x}(r,\x,\y)=r^3c_{t\x}(\x,\y)+\mathcal{O}(r^4)\,,\nonumber\\
&Q_{t\y}(r,\x,\y)=r^3c_{t\y}(\x,\y)+\mathcal{O}(r^4)\,,\quad Q_{r\x}(r,\x,\y)=r^4c_{r\x}(\x,\y)+\mathcal{O}(r^4 \,lnr)\,,\nonumber\\
&Q_{r\y}(r,\x,\y)=r^4c_{r\y}(\x,\y)+\mathcal{O}(r^4 \,lnr)\,,\nonumber\\
&a_t(r,\x,\y)=\mu+r c_t(\x,\y)+\mathcal{O}(r^2)\,,\quad a_r(r,\x,\y)=r^3 c_r(\x,\y)+\mathcal{O}(r^3 \,lnr)\,,\nonumber\\
&a_{\x}(r,\x,\y)=r c_{\x}(\x,\y)+\mathcal{O}(r^2)\,,\quad a_{\y}(r,\x,\y)=r c_{\y}(\x,\y)+\mathcal{O}(r^2)\,,\nonumber\\
&\phi_1(r,\x,\y)=r c_1(\x,\y)+\mathcal{O}(r^2)\,,\quad \phi_2(r,\x,\y)=r c_2(\x,\y)+\mathcal{O}(r^2)\,.
\end{align}
The appearance of the function $c_{tt}$ in the expansion for $Q$ is associated with the fact that the
stress tensor is traceless.
We have sixteen functions of $\x,\y$ which are fixed in the numerical integration. 
The functions $g_1$ and $g_2$ have been
seen in similar DeTurck constructions before \cite{Donos:2014yya,Donos:2015eew}, and can be removed by a gauge transformation
when the modifications to the equations of motion vanish, $\xi^\mu=\varphi=0$. Similarly, $c_{tr}$, $c_{r\x}$, $c_{r\y}$ and $c_r$ can be removed by gauge transformations, if desired; they don't appear in the expressions for physical quantities in the boundary theory.

In order to calculate the one point functions we need to supplement the bulk action \eqref{eq:action} with
the boundary action given in \eqref{ctermp} (in the Euclidean frame). The 
expectation value of the stress tensor, $T^\mu{}_{\nu}$, can be obtained from the expression 
\begin{align}\label{stressy}
T^\mu{}_{\nu}\equiv& \frac{1}{\tilde L_x \tilde L_y} \,\lim_{r\to 0}\,r\, \sqrt{-g}\,[-2K^\mu{}_{\nu}+\delta^\mu{}_{\nu}(2K-4)]\,,
\end{align}
where $g$ refers to the bulk metric, and we find
\begin{align}\label{stress}
&T^t{}_t=-\frac{B^2}{2 r_+}-\frac{\mu^2 r_+}{2}-2 r_+^3+\frac{3}{8} r_+^3c_{tt}\,,\nonumber\\
&T^{\x}{}_{\x}=\frac{B^2}{4 r_+}+\frac{\mu^2 r_+}{4}+r_+^3-\frac{3}{16} r_+^3c_{tt}+\frac{3}{8} r_+^3 (1+v^2) c_W\,,\nonumber\\
&T^{\y}{}_{\y}=\frac{B^2}{4 r_+}+\frac{\mu^2 r_+}{4}+r_+^3-\frac{3}{16} r_+^3c_{tt}-\frac{3}{8} r_+^3 (1+v^2) c_W\,,\nonumber\\
&T^{\x}{}_{\y}=\frac{3\tilde L_y r_+^3}{8 \tilde{L}_x \sqrt{1+v^2}}[v(1+v^2) c_W+c_R]\,,\quad
T^{\y}{}_{\x}=-\frac{3\, r_+^3 \tilde{L}_x}{8 \tilde{L}_y\,\sqrt{1+v^2} }(v(1+v^2) c_W-c_R)\,,\nn
%&T^t{}_{\x}=\frac{3}{8}\tilde L_x r_+^3c_{t \x}\,,\qquad
%T^t{}_{\y}=\frac{3}{8}\tilde L_y r_+^3c_{t \y}\,,\nn
&T^{\x}{}_t=-\frac{3\, r_+^3}{8 \tilde{L}_x}(\sqrt{1+v^2}  c_{t\x}) -v c_{t\y})\,,\quad
T^{\y}{}_t=\frac{3\, r_+^3}{8 \tilde{L}_y}( v c_{t\x}-\sqrt{1+v^2}  c_{t\y})\,.
\end{align}
Notice that these expansions imply the Ward identity $T^t{}_t+T^{\x}{}_{\x}+T^{\y}{}_{\y}=0$.
Furthermore, the expectation values of the abelian current density are given by
 \begin{align}
 J^a= \frac{1}{\tilde L_x \tilde L_y} \,\lim_{r\to 0}\sqrt{-g} \,F^{ar}\,,
 \end{align}
and we find 
\begin{align}\label{jayapp}
&J^t=r_+(\mu-\frac{1}{2} c_t)\,,\nonumber\\
&J^x=\frac{r_+}{2 L_x}(\sqrt{1+v^2}c_{\x}- v c_{\y})\,,\nonumber\\
&J^y=\frac{r_+}{2 L_y}(\sqrt{1+v^2}c_{\y}- v c_{\x})\,.
\end{align}
The expectation value for the operator, $\mathcal{O}_\psi$, dual to the complex scalar, $\psi=\psi_1+i\psi_2$, 
can also be determined\footnote{If the asymptotic metric is $ds^2\to \epsilon^{-2}(d\epsilon^2-dt^2+dx^2+dy^2)$ as $\epsilon\to 0$,
then in the expansion $\psi=\epsilon\psi_I+\epsilon^2\psi_{II}+\dots$ we identify $\langle \mathcal{O}_\psi \rangle$ with $\psi_{II}$.}.
Writing $\langle |\mathcal{O}_{\psi}|\rangle$ for the modulus of $\langle \mathcal{O}_\psi\rangle$ we find
\begin{align}
\langle|\mathcal{O}_{\psi}|\rangle = \frac{ r_+^2}{2} \sqrt {c_1^2+c_2^2}\,.
\end{align}
Similarly the sine of the phase of $\langle \mathcal{O}_\psi\rangle$ is given by $c_1/\sqrt{c_1^2+c_2^2}$ and we find
that the phase winds once around the core of each vortex in the solutions we have constructed, i.e. $n=1$.

At fixed $\mu,B, T$, within numerical precision we find that  
for all of the black holes we have constructed we have
\begin{align}
\bar c_{t\x}=\bar c_{t\y}=0\,,\qquad \bar c_{\x}=\bar c_{\y}=0\,,
\end{align}
and this corresponds to the fact that in thermal equilibrium the black holes
have vanishing current fluxes:
\begin{align}
\bar T^i{}_t&=\bar J^i=0
\end{align}
Furthermore, for the thermodynamically preferred black holes,
identified by varying the free energy density $w$ with respect to $v$ and $k$ (defined in \eqref{kdef}),
within numerical precision we find for all black hole solutions
\begin{align}
\bar c_R=\bar c_W=0\,.
\end{align}
This corresponds to the conditions
\begin{align}
\bar T^i{}_j&=p\delta^i{}_j,\qquad -\frac{1}{2}\bar T^t{}_t=p\,.
\end{align}

\section{Numerical Convergence}\label{appendixB}
To solve the system of PDEs discussed in section \ref{sec:numerics}, with appropriate boundary conditions, we discretised the computational domain $[0,1]\times [0,1]\times [0,1]$ using $N_r, N_x, N_y$ points, respectively. We approximated the derivatives of the fields on these points using an interpolation method: since the $\x$ coordinate is periodic, we used a Fourier spectral method, while for $r$ and $\y$ we used the Chebyshev spectral method (recall that we have to implement the quasi-periodic boundary conditions under shifts of $\y \to \y+1$ as in \eqref{eq:bcscalars}). 
The resulting algebraic system of equations was solved iteratively using the Newton-Raphson method.
For all plots given in the paper we have taken $N_r=N_y=50$ and $N_x=46$, unless stated otherwise.

\begin{figure}[h!]
\centering
\includegraphics[width=0.46\linewidth]{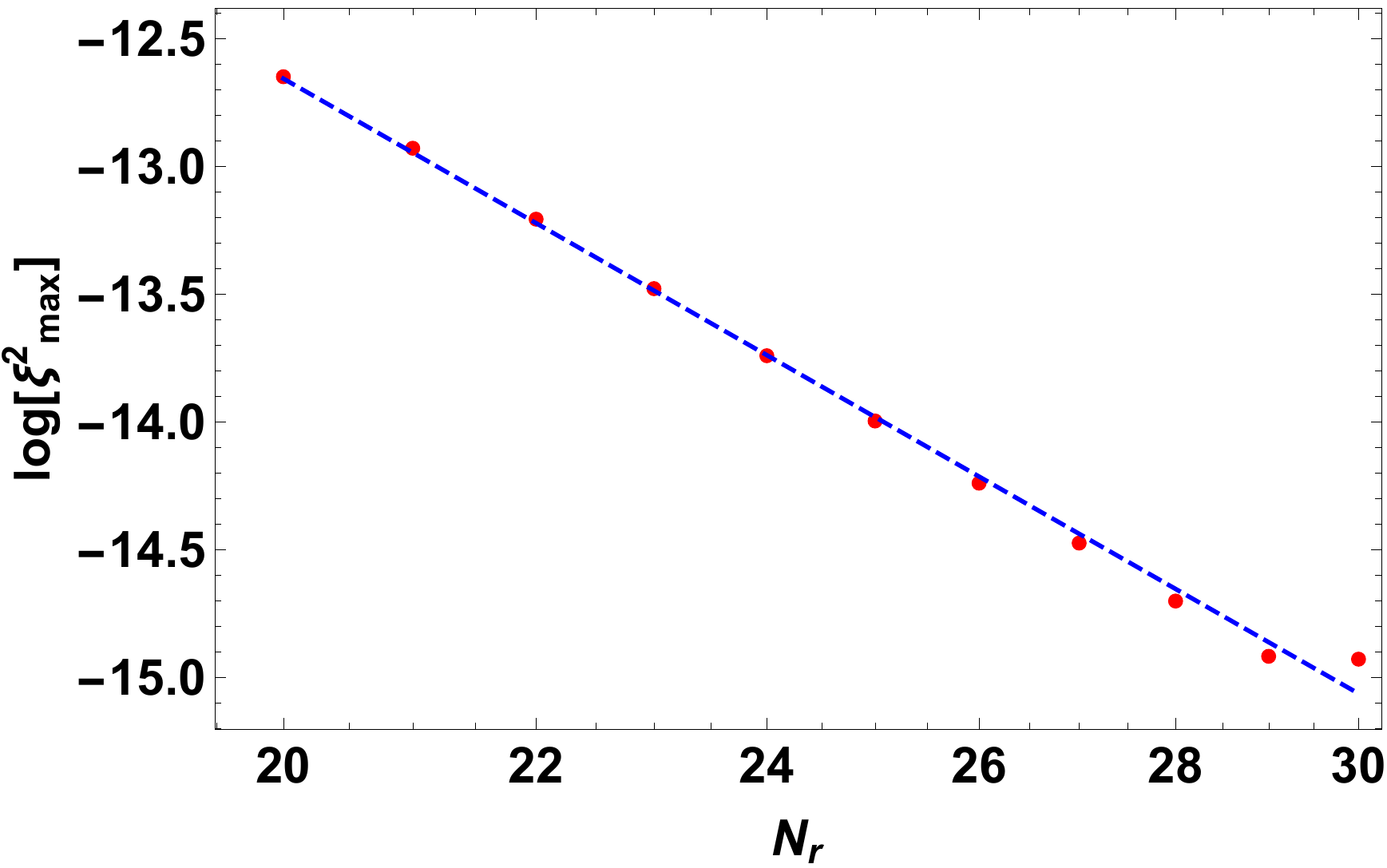}
\caption{Plot of $\xi^2_{\max}$, the maximum value of $\xi^{2}$,  as a function of the radial resolution $N_r$, 
for the preferred triangular vortex lattice black holes (lowest Landau level) at $T/\mu=0.08, B=0.01$. Here $N_x=60, N_y=50$.}
\label{fig:tests}
\end{figure}

We now discuss the convergence of our code with increasing lattice resolution, especially with the number of point $N_{r}$ that we take in the radial direction. One reason this is important is because we have no analytic proof that the only possible solutions of the DeTurck system of equations which we solve actually reduce to Einstein's equations when we impose our boundary conditions. We therefore need to check that $\xi_{\mu}$ becomes trivial everywhere in our domain in the continuum limit. This is unlike the case of the auxiliary field $\varphi$ which we introduced in \eqref{moddtwo} in order to fix gauge invariance. In fact, we can show that $\varphi$ must be trivial and it can therefore serve as a check for the convergence of our code. In figures \ref{fig:tests} we have plotted $\xi^2_{\max}$, the maximum value of $\xi^{2}$ that we find in our computational domain. Fitting that to a power law we find the behaviour $\sim N_r^{-13.7}$. We found it reassuring that in all our solutions we had $|\varphi|<10^{-8}$. However, this error is close to our numerical precision and so we are not able to check its convergence properties.

A second reason that the details of the convergence of our code is important
concerns the non-analytic terms related to $g_{1}$ and $g_{2}$ which appear in the near conformal boundary expansion \eqref{eq:UVexp}. In general, such terms turn the exponential convergence of spectral methods to power law with 
respect\footnote{On the one hand this means that the spectral methods which we use in the radial direction become inefficient at high resolutions. On the other hand, however, in terms of memory usage, we have found them to be very efficient in
obtaining sufficiently accurate solutions. Indeed it would be significantly
more demanding in resources in order to achieve
similar accuracy
using finite difference methods in the radial direction.
It is also worth noting that the precise power law behaviour which we find is for a specific choice of reference
metric in \eqref{moddone}, namely the AdS-RN black hole solution. It would be interesting to explore whether other choices leads to improved convergence.}
 to the radial direction
(as we saw above for the behaviour of $\xi^2$). For the purposes of our checks, we have found useful to consider the quantities,
\begin{align}
c^{(1)}&=\left.\frac{2}{3}\left(\frac{Q_{tt}-Q}{r}\right)''\right|_{r=0}\,,\label{eq:ctt1}\\
c^{(2)}&=\left.\left(\frac{Q_{tt}-1}{r}\right)''\right|_{r=0}\,.\label{eq:ctt2}
\end{align}
Using the expansion \eqref{eq:UVexp}, we can easily see that in the continuum we must have $c_{tt}=c^{(1)}=c^{(2)}$. Notice that this quantity is particularly important since it enters the expression for the free energy of the system via
\eqref{stress} and \eqref{eq:OSactionp}. From the expansions \eqref{eq:UVexp}, we see that the leading non-analytic power will drop out from $c^{(1)}$ and we therefore expect it to have a better convergence rate than $c^{(2)}$.

\begin{figure}[h!]
\centering
\includegraphics[width=0.46\linewidth]{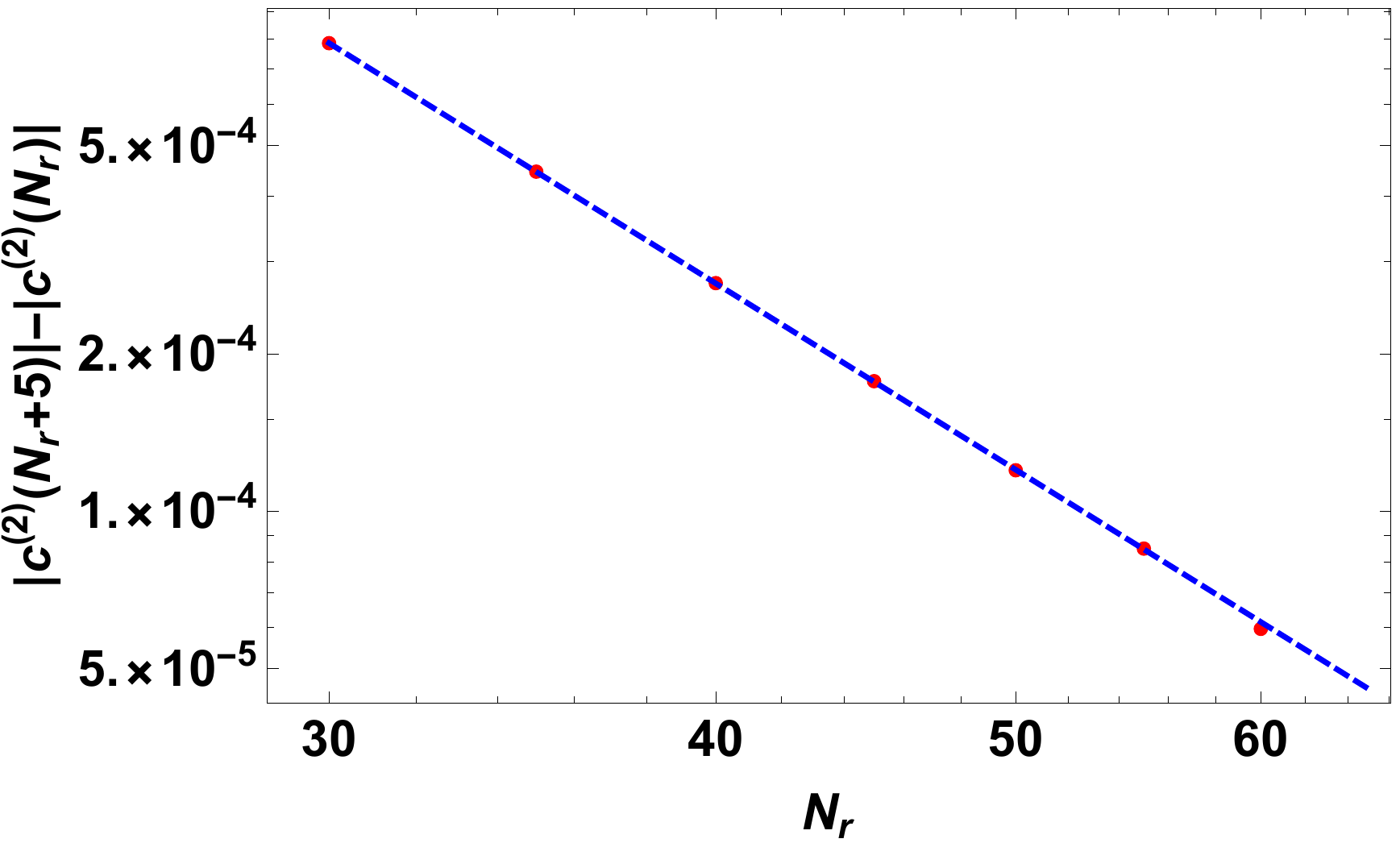}
\caption{Plot of the behaviour of $c^{(2)}$ as a function of the radial resolution $N_r$,
for the preferred triangular vortex lattice black holes (lowest Landau level) at $T=0.08, B=0.01$.
Here $N_x=31, N_y=30$.}
\label{fig:ctt}
\end{figure}

In figure \ref{fig:ctt} we plot $c^{(2)}(N_r+5)-c^{(2)}(N_r)$ as a function of the points in the radial direction, $N_r$.  We see that this quantity has power law convergence to zero, with behaviour $\sim N_r^{-3.7}$. Our expectation is that 
$c^{(1)}(N_r+5)-c^{(1)}(N_r)$ would converge with a smaller power. However, we were not able to confirm this since, due the absence of the leading non-analytic terms and faster convergence, it becomes smaller than our numerical precision quite quickly as we increase $N_{r}$, before it takes the asymptotic form of a power law\footnote{For a lattice with $N_{x}=36$ and $N_{y}=36$ we found, for example, that $c^{(1)}(N_r+1)-c^{(1)}(N_r)\sim 10^{-9}$ for $N_{r}\geq 30$.}.
It is worth highlighting that these tests for $c^{(1)}$ $c^{(2)}$ are associated with the DeTurck equations we are solving and are not related to the fact that we are not solving Einstein's equations. On the conformal boundary we don't impose the equations of motion but our boundary conditions. We therefore need to check whether the equations of motion are asymptotically satisfied by numerically checking the series expansion \eqref{eq:UVexp} of our functions.

The main error in the various tests we have discussed arises from different places in the bulk. For figure \ref{fig:ctt} the main error comes from the near conformal boundary region, as one might expect from the discussion above. On the other hand, the error for the plot in figure \ref{fig:tests} arises from a more global test and, in fact, mainly arises from the near horizon region where $\xi^{2}$ and $|\varphi |$ take their maximum values.

We have implemented our numerical method in C++ using double precision numbers and class data structures. The code is fully parallelised in shared and distributed memory using a combination of MPI and OpenMP. After fixing a discretisation scheme, the problem reduces to solving a set of non linear algebraic equations for the values of our functions on the grid described above. This is done using the Newton-Raphson method where one starts with an initial guess for the unknown functions at each lattice point and then iteratively corrects it in order to obtain functions that solve the PDEs to a better and better approximation. This boils down to solving a linear system at each step in Newton's method. To do this we use the PETSc library with a block-ILU(0) preconditioner in combination with a GMRES iterative method. 

%\bibliographystyle{utphys}
%\bibliography{helical}{}

\providecommand{\href}[2]{#2}\begingroup\raggedright\endgroup

\end{document}